\documentclass[review,9pt]{elsarticle}

\usepackage{lineno,hyperref}
\usepackage{multirow}
\usepackage{color}
\usepackage{subfig}
\usepackage{soul}
\usepackage{breqn}
\usepackage[table,xcdraw]{xcolor}
\usepackage{stackengine}
\usepackage{tabularx}
\usepackage{todonotes}
\usepackage{booktabs}
\usepackage{xcolor}
\usepackage{soul}
\usepackage{todonotes}
\usepackage{amsmath,amsthm,amssymb}
\usepackage{graphicx}
\usepackage{adjustbox}
\usepackage{rotating}
\usepackage{hhline}
\usepackage{verbatim} 
\modulolinenumbers[5]

\journal{Journal of Applied Acoustics}

\usepackage{todonotes}

\begin{document}

\begin{frontmatter}

\title{{\textcolor{black}{From Environmental Sound Representation to Robustness of 2D CNN Models Against Adversarial Attacks}}\footnote{{Supplementary materials and source codes are available at} \href{https://github.com/EsmaeilpourMohammad/From-Sound-Representation-to-Model-Robustness.git}{{this GitHub repository.}}}}



 \author{Mohammad Esmaeilpour\corref{}}
 \ead{mohammad.esmaeilpour.1@ens.etsmtl.ca} 
 \author{Patrick Cardinal}
 \ead{patrick.cardinal@etsmtl.ca} 
  \author{Alessandro Lameiras Koerich}
 \ead{alessandro.koerich@etsmtl.ca} 
 \address{D\'{e}partement de G\'{e}nie Logiciel et des TI\\ \'{E}cole de Technologie Sup\'{e}rieure (\'{E}TS)\\
Universit\'{e} du Qu\'{e}bec\\
1100 Notre-Dame W, Montr\'{e}al, H3C 1K3, Qu\'{e}bec, Canada}




\begin{abstract}
\textcolor{black}{This paper investigates the impact of different standard environmental sound representations (spectrograms) on the recognition performance and adversarial attack robustness of a victim residual convolutional neural network, namely ResNet-18. Our main motivation for focusing on such a front-end classifier rather than other complex architectures is balancing recognition accuracy and the total number of training parameters. Herein, we measure the impact of different settings required for generating more informative Mel-frequency cepstral coefficient (MFCC), short-time Fourier transform (STFT), and discrete wavelet transform (DWT) representations on our front-end model. This measurement involves comparing the classification performance over the adversarial robustness. We demonstrate an inverse relationship between recognition accuracy and model robustness against six benchmarking attack algorithms on the balance of average budgets allocated by the adversary and the attack cost. Moreover, our experimental results have shown that while the ResNet-18 model trained on DWT spectrograms achieves a high recognition accuracy, attacking this model is relatively more costly for the adversary than other 2D representations. We also report some results on different convolutional neural network architectures such as ResNet-34, ResNet-56, AlexNet, and GoogLeNet, SB-CNN, and LSTM-based.}
\end{abstract}


\end{frontmatter}


\section{Introduction}
Developing reliable sound recognition algorithms for real-life applications has always been a significant challenge for the signal processing community~\cite{marchegiani2017leveraging,salamon2017scaper,radhakrishnan2005audio}. For analyzing the surrounding scene either for surveillance \cite{valenzise2007scream} or multimedia sensor networks \cite{steele2013sensor}, there is a constant need to understand environmental events. Raised by these concerns, several unsupervised \cite{salamon2015unsupervised} and supervised \cite{salamon2017deep} algorithms have been devised for classifying environmental sounds. 
During the last decades, there has been increasing attention toward developing automatic environmental sound classifiers. Presumably, this is due to its vast applications in smart acoustic sensor network development \cite{mydlarz2017implementation}, surveillance scene monitoring \cite{radhakrishnan2005audio,cristani2004line}, IoT-based noise reduction \cite{shah2019iot}, smart city safety \cite{ciaburro2020improving,shah2018audio}, and context-aware computing \cite{chandrakala2021multi,chu2009environmental,toffa2020environmental}. Towards developing reliable classification algorithms for such tasks, the impact of adversarial attacks on the deep learning (DL) classifiers trained on environmental sounds should be investigated.
In other words, developing reliable environmental sound classifiers requires the study of adversarial attacks in greater detail to account for the impact of such attacks on different sound representations. This is our main motivation for setting the framework of this paper to environmental sounds encompassing a broad spectrum of urban sounds. 


With the proliferation of DL algorithms during the last decade for image-related tasks, many publications on audio representations (spectrograms) have been released \cite{chu2008environmental,chu2009environmental,huzaifah2017comparison,dave2013feature,tsai2015novel}. The DL architectures primarily developed for computer vision applications have been well adapted for sound recognition tasks with recognition accuracy (following the basic statistical definition in \cite{pedregosa2011scikit}) competitive to human understanding. However, such algorithms require large amounts of training data. In response, many low-level data augmentation approaches have been introduced to allow an appropriate training of DL models and improve their performance on sound-related tasks \cite{salamon2017deep}. These approaches apply directly to audio waveforms affecting low-level sampled data points of the audio signal, which may not necessarily improve the performance of the front-end classification models \cite{esmaeilpour2019unsupervised}. High-level data augmentation approaches have been developed to tackle this problem, which are particularly useful for audio representations \cite{kaneko2017generative, mathur2019mic2mic}. Experimental results on a variety of environmental sound datasets attest considerable positive impact of high-level data augmentation on overall performance of DL classifiers (e.g., AlexNet \cite{krizhevsky2012imagenet}, GoogLeNet \cite{szegedy2015going}, etc.) \cite{esmaeilpour2019unsupervised}.

Unfortunately, recent studies have demonstrated the vulnerability of these convolutional neural networks (ConvNets) trained on 2D representations of audio signals against adversarial attacks \cite{esmaeilpour2019robust}. They have shown that crafted adversarial examples are transferable among dense ConvNets and support vector machines (SVM). That poses potential harm for sound recognition systems, especially when the highest recognition accuracy has been reported on 2D representations over raw 1D audio signals \cite{boddapati2017classifying}. This threat negatively affects the reliability of DL models designed for applications based on sound classification, particularly IoT-related tasks in an environmental setting. \cite{zamil2019multimedia}.

Toward proposing reliable classifiers, there have been some debates and case studies on the link between intrusion of adversarial examples and loss functions for some victim classifiers \cite{carlini2017towards}. It has been shown that the integration of more convex loss functions in the victim model (or in the surrogate counterpart) might increase the chance of crafting more potent adversarial examples \cite{carlini2017towards}. However, it might also depend on some other key factors such as the properties of the classifier, input sample distribution, adversarial setups, etc. To study other potential links, we evaluate the robustness and the transferability of some state-of-the-art ConvNets against adversarial attacks trained on different 2D environmental sound representations. Our primary front-end ConvNet is ResNet-18 architecture because of its superior recognition performance compared to other ConvNet architectures. We discuss this in Section~\ref{classification_model_Sec} and briefly report our findings on different dense architectures such as GoogLeNet and AlexNet in Section~\ref{discuss:sec}.

\textcolor{black}{The main novelty in this paper is investigating the response of a state-of-the-art classifier (i.e., ResNet-18)} to different 2D representations in terms of both the recognition accuracy and the robustness against adversarial attacks. This helps to yield more reliable classifiers without running any costly adversarial defense algorithm. More specifically, we make the following contributions:
\textcolor{black}{
\begin{enumerate}[(i)]
   \item We show that ResNet-18 models achieve higher recognition accuracy on the DWT representation than STFT and MFCC averaged over different spectrogram settings for three comprehensive environmental sound datasets.
   {This result can be helpful for researchers who would like to develop practical audio classification algorithms based on spectral features. Furthermore, considering that DWT outperforms the two other audio representations (for the environmental sound analysis) contributes to building a more robust classifier.}
    \item We identify major spectrogram settings which considerably affect the cost of attack (the number of required gradient computations) averaged over budgets. 
    {One of the open problems and the real-life challenges in machine learning research is developing reliable defense algorithms against varieties of adversarial attacks \cite{ozdag2018adversarial}. Thus, by choosing the proper settings during the spectrogram production, it is possible to increase the attack cost for the adversary and, to some extent, protect victim classifiers.} 
    \item We characterize the existence of an inverse relation between recognition accuracy and robustness of the victim models (i.e., ResNet-18) against six strong targeted and non-targeted benchmarking adversarial attacks. On average, models with higher recognition accuracies undergo higher fooling rates. 
    {This contribution can be beneficial for researchers who would like to make a reliable trade-off between recognition accuracy and the robustness of the classifiers against adversarial attacks \cite{esmaeilpour2019robust}. However, making this trade-off can be challenging and time-consuming for some datasets. Hence, bearing the inverse relationship mentioned above (between recognition accuracy and adversarial robustness) might contribute to achieving a more steady trade-off.} 
    \item We demonstrate that compared to DWT and STFT, the MFCC has a relatively lower adversarial transferability ratio among three advanced DL architectures, namely ResNet-18, AlexNet, and GoogLeNet.
    {Unfortunately, it has been proven and demonstrated that adversarial examples are transferable among different data-driven classifiers \cite{papernot2018characterizing}. Therefore, toward developing a classifier with a lower transferability ratio, our technical recommendation would be to exploit the MFCC representation.} 
\end{enumerate}
}

The rest of the paper is organized as follows. In Section~\ref{sec:adv}, we briefly review some strong adversarial attacks for audio representations. Then, explanations on different audio representations that have been used in the experiments are summarized in Section~\ref{audio_rep_section}. Next, experimental results and associated discussions are presented in Sections~\ref{sec:exp} and~\ref{discuss:sec}, respectively. Finally, the conclusions and perspectives of future work are presented in the last section. 

\section{Adversarial Attacks}
\label{sec:adv}
Assuming we have a sound recognition system that employs a classifier trained on legitimate spectrograms. In the following, we explain how crafted adversarial spectrograms can pose security concerns for this system.
\begin{itemize}
    \item \textbf{White-box scenario:} The adversary has full access to the entire system details, including audio dataset, classifier architecture, potential tuning parameters, required hyperparameters, and complete weight vectors. Therefore, the adversary can easily feed adversarial spectrograms to the model and fool it toward any incorrect target label.
    \item \textbf{Black-box scenario:} The adversary does not have access to the system mentioned above details. Thus, the adversary can only input a 1D signal to the system and receive a predicted label. In this scenario, the adversary can reconstruct an audio signal from an adversarial spectrogram (with or without a surrogate model) and feed it to the system. Since the model is trained on spectrograms, the system first converts the input audio into a spectrogram that embeds the adversarial perturbation. This reconstruction does not pose a technical difficulty since spectrogram and 1D signal are dual, and there are plenty of straightforward approaches for reconstructing one from another. However, this spectrogram can also fool the model toward any wrong label defined by the adversary (see a relevant study in \cite{koerich2020cross}).
\end{itemize}

\subsection{Adversarial Attack For Environmental Sound Classifiers}
In practice, adversarial attacks exist both for 1D signals~\cite{li2020advpulse} and their associated 2D representations~\cite{esmaeilpour2019robust}. This paper focuses on the latter because, for some decades, spectrograms (generated from MFCC, STFT, DWT) have been relatively standard representations for different audio and speech processing tasks, particularly classification. Besides, spectrogram and 1D signal are duals (bijectively convertible), and the highest recognition accuracy on the benchmarking environmental sound datasets have been reported for models trained on the 2D representations~\cite{boddapati2017classifying}. 

Technically, an adversarial attack can be formulated as an optimization problem toward achieving a minimal perturbation parameter $\delta$ as stated in Eq.~\ref{general_adv_formula} \cite{szegedy2013intriguing}.
\begin{equation}
 \min_{\delta} \quad f^{*}(\bold{x}+\delta) \neq f^{*}(\bold{x})
 \label{general_adv_formula}
\end{equation}
\noindent where $\bold{x}$ and $f^{*}$ denote a legitimate random spectrogram and the post-activation function of the victim classifier, respectively. The value for $\delta$ should be as small as possible to not being perceivable by humans. Many attack algorithms that satisfy such an imperceptibility constraint have been proposed in white and black-box scenarios. In this paper, we briefly go over six strong targeted and non-targeted adversarial attacks, which are well adapted to sound recognition models trained on audio representations \cite{esmaeilpour2019robust}. We use the average fooling rate of these attacks, a standard metric for assessing the robustness of victim ConvNets trained on different audio representations.

\subsection{Limited-Memory Broyden-Fletcher-Goldfarb-Shanno (L-BFGS)}
Szegedy \textit{et al.} \cite{szegedy2013intriguing} argue that the viability of fooling deep neural networks with fake examples is due to their extremely low probability because such examples are rarely seen in a given dataset. That could be understood as the pitfall of deep networks in low generalizability to unseen but very similar samples. However, they propose an optimization algorithm to mislead finely trained DL models, based on Eq.~\ref{lbfgs}:
\begin{equation}
 \min_{\bold{x}^\prime} c\left \| \delta \right \|_{2} + J_{\bold{w}}(\bold{x}^\prime,l^\prime)
 \label{lbfgs}
\end{equation}
\noindent where $c$ is a positive scaling factor achievable by the line search strategy, $\bold{x}^\prime$ denotes the associated crafted adversarial example, $l^\prime$ refers to its target label, and $J_{\bold{w}}$ denotes the loss function for updating weights ($\bold{w}$). There are various choices for this function, such as cross-entropy loss or any other surrogate function. The solution to this optimization problem is quite costly, and it has been proposed to use 
the L-BFGS optimizer, subject to $0\leq\bold{x}^\prime\leq M$ where $M$ refers to the maximum possible intensity in a spectrogram. This attack is the baseline for the adversarial algorithms that are subsequently presented. 

\subsection{Fast Gradient Sign Method (FGSM)}
Goodfellow \textit{et al.} \cite{goodfellow2014explaining} explain the existence of adversarial examples with linear nature of deep neural networks, even those with super-dense hidden layers. Toward this claim, they proposed a fast optimization algorithm based on Eq.~\ref{fgsm}:
\begin{equation}
 \bold{x}^\prime\leftarrow \bold{x}+\delta \cdot \mathrm{sign}(\nabla_{\bold{x}}J(\bold{x},l))
 \label{fgsm}
\end{equation}
\noindent where $\delta$ is a small constant for controlling the applied perturbation to the legitimate sample $\bold{x}$. Different choices of $\ell_{p}$ norms can be integrated into the FGSM attack, and the adversary should make a trade-off between high similarities and a large enough perturbation to be able to fool 
a model. The formulation of Eq.~\ref{fgsm} for $\ell_{2}$ norm is shown in Eq.~\ref{fgsm2}.
\begin{equation}
 \bold{x}{}'\leftarrow \bold{x}+ \delta \frac{\nabla_{\bold{x}}J(\bold{x},l)}{\left \| \nabla_{\bold{x}}J(\bold{x},l) \right \|}
 \label{fgsm2}
\end{equation}
\noindent where for satisfying the constraint $\bold{x}^\prime\in[0,M]$, the resulting adversarial spectrogram should be clipped or truncated. This white-box adversarial attack is targeted toward a pre-defined wrong label by the adversary in a one-shot scenario.

\subsection{Basic Iterative Method (BIM)}
This non-targeted adversarial attack \cite{kurakin2016adversarial} is, in fact, the iterative version of the FGSM optimization algorithm, which crafts and positions potential adversarial examples ideally outside of legitimate subspaces via optimizing Eq.~\ref{bim} for $\delta$:
\begin{equation}
 \bold{x}^\prime_{n+1} \leftarrow \mathrm{clip}_{\bold{x},\delta}\begin{Bmatrix}
\bold{x}^\prime_{n} + \delta \cdot \mathrm{sign} (\nabla_{\bold{x}}J(\bold{x}_{n},l))
\end{Bmatrix}
\label{bim}
\end{equation}
\noindent where $\mathrm{clip}$ is a function for keeping generated examples within the range $[\bold{x}-\delta, \bold{x}+ \delta]$ as defined in Eq.~\ref{clip}.
\begin{equation}
\min \begin{Bmatrix}
M, \bold{x}+\delta, \max\{0,\bold{x}-\delta,\bold{x}^\prime\}
\end{Bmatrix}
\label{clip}
\end{equation}
\noindent where $M$=255 for 8-bit RGB visualization of spectrograms. 

There are two implementations for this optimization algorithm either by optimizing up to reach the first adversarial example (BIM-a) or continuing optimizing to a predefined number of iterations (BIM-b). The latter usually generates stronger adversarial examples, though it is more costly since it usually requires more callbacks. Both BIM attacks are iterative and white-box algorithms minimizing Eq.~\ref{bim} for optimal perturbation $\delta$ measured by $\ell_{\infty}$ norm.

\subsection{Jacobian-based Saliency Map
Attack (JSMA)}
Similar to the FGSM attack, this algorithm also uses gradient information for perturbing the input taking advantage of a greedy approach \cite{papernot2016limitations}. This attack is targeted toward a pre-defined wrong label ($l^\prime$). In fact, it optimizes for $\arg\min_{\delta_{\bold{x}}}\left \| \delta_{\bold{x}} \right \|$ subject to $f^{*}(\bold{x}+\delta_{\bold{x}})=l^\prime$ (optimizing with $\ell_{0}$). There are three steps in developing JSMA adversarial examples. First, computing the derivative of the victim model as Eq.~\ref{jsma_jacob}.
\begin{equation}
 \nabla f(\bold{x})=\frac{\partial f_{j}(\bold{x})}{\partial x_{i}}
 \label{jsma_jacob}
\end{equation}
\noindent where $x_{i}$ denotes pixels intensities. Second, a saliency map should be computed to detect the least effective pixel values for perturbation according to the desired outputs of the model. Specifically, the saliency map for pixels in cases where $\partial f_{l}(\bold{x})/\partial \bold{x}_{i} < 0$ or $\sum_{j\neq l}\partial f_{j}(\bold{x})/\partial \bold{x}_{i}> 0$ should be set to zero since there are detectable variations, otherwise:
\begin{equation}
 S_{map}(\bold{x},l^\prime)[i] = \frac{\partial f_{l}(\bold{x})}{\partial \bold{x}_{i}}\left | \sum_{j\neq l^\prime}\frac{\partial f_{j}(\bold{x})}{\partial \bold{x}_{i}} \right |
 \label{jsma_case2}
\end{equation}
\noindent where $S_{map}$ denotes the saliency map for every given spectrogram $\bold{x}_{i}$ and target label $l^\prime_{i}$. The last step of the JSMA is applying the perturbation on the original input according to the achieved map.

\subsection{Carlini and Wagner Attack (CWA)}
This is an iterative and white-box adversarial algorithm \cite{carlini2017towards}, which can use three types of distance metrics: $\ell_{0}$, $\ell_{\infty}$, and $\ell_{2}$ norms. 
This paper focuses on the latter distance measure making the algorithm very strong even against the distillation network. The optimization problem in this attack is given by Eq.~\ref{cwa_minim}.
\begin{equation}
 \min_{\delta}\left \| \bold{x}^\prime- \bold{x} \right \|_{2}^{2}+cf(\bold{x}^\prime)
 \label{cwa_minim}
\end{equation}
\noindent where $c$ is a constant value as explained in Eq.~\ref{lbfgs}. Assuming the target class is $l^\prime$ and $G(\bold{x}^\prime)_{i}$ denotes the logits of the trained model $f$ before softmax activation corresponding to the $i$-th class, then:

\begin{equation}
 f(\bold{x}^\prime)=\max \left\{ \max_{i\neq l^\prime}\left \{ G(\bold{x}^\prime)_{i}\right \} - G(\bold{x}^\prime)_{l^\prime}, -\kappa \right \}
 \label{cwa_func}
\end{equation}
\noindent where $\kappa$ is a tunable confidence parameter for increasing misclassification confidence toward label $l^\prime$, the actual adversarial example is given by Eq.~\ref{cwa_advEx}.
\begin{equation}
 \bold{x}'=\frac{1}{2}\left [ \tanh(\mathrm{arctanh}(\bold{x})+\delta)+1 \right ]
 \label{cwa_advEx}
\end{equation}
\noindent where the $\tanh$ activation function is used in replacement of box-constraint optimization. For non-targeted attacks, Eq.~\ref{cwa_func} should be updated as: 
\begin{equation}
f(\bold{x}^\prime) = \max \left\{ G(\bold{x}^\prime)_{l}-\max_{i\neq l}\left \{ G(\bold{x}^\prime)_{i} \right \} ,-\kappa\right\}
\end{equation}
\subsection{DeepFool Adversarial Attack}
Moosavi-Dezfooli \textit{et al.} \cite{moosavi2016deepfool} proposed a white-box algorithm for finding the most optimal perturbation for redirecting the position of a legitimate sample toward a pre-defined target label using linear approximation. The optimization problem for achieving optimal $\delta$ is given by Eq.~\ref{deepfool_eq1}.
\begin{equation}
 \arg \min \left \| \delta \right \|_{2} \quad \mathrm{s.t.} \quad \mathrm{sign}(f(\bold{x}^\prime))\neq \mathrm{sign}(f(\bold{x}))
 \label{deepfool_eq1}
\end{equation}
\noindent where $\delta = -f(\bold{x})\bold{w}/\left \| \bold{w} \right \|_{2}^{2}$. DeepFool can also be modified to a non-targeted attack optimizing for hyperplanes of the victim model. In this paper, we implement targeted DeepFool attack and averaged over available labels measuring over $\ell_{2}$ and $\ell_{\infty}$. In practice, this scenario is not only faster but also more destructive than BIMs. 

A summary of the aforementioned adversarial attacks with their properties is provided in Table~\ref{table:propertiesCompbench}. This table briefly explains the advantages of every attack over another. The following section gives a brief overview of common 2D representations of audio signals using time-frequency transformations. Finally, we carry out our adversarial experiments on the transformed audio signals (spectrograms).

\begin{table*}[t]
\centering
\scriptsize
\caption{Properties comparison of the benchmarking adversarial attacks.}
\begin{tabular}{|c||c|c|c|c|c|}
\hline
Attack                                                      & \begin{tabular}[c]{@{}c@{}} {Key positive}\\ {feature}\end{tabular}             & \begin{tabular}[c]{@{}c@{}} {Potential}\\ {Compromise}\end{tabular}                 & \begin{tabular}[c]{@{}c@{}} {Targeted or}\\ {non-targeted}\end{tabular} & \begin{tabular}[c]{@{}c@{}} {Similarity}\\ {metric}\end{tabular} & \begin{tabular}[c]{@{}c@{}}{Optimization}\\ {policy}\end{tabular} \\ \hline \hline
{FGSM}                                                        & \begin{tabular}[c]{@{}c@{}}{Very fast and}\\ {flexible}\end{tabular}           & \begin{tabular}[c]{@{}c@{}}{Easily}\\ {detectable}\end{tabular}                    & {Targeted}                                                           & $l_{\infty}$                                                & {One-shot}                                                      \\ \hline
{DeepFool}                                                    & \begin{tabular}[c]{@{}c@{}}{Linear}\\ {formulation}\end{tabular}               & \begin{tabular}[c]{@{}c@{}}{Costly in}\\ {optimization}\end{tabular}               & {Non-Targeted}                                                       & $l_{2}$, $l_{\infty}$                                       & {Iterative}                                                     \\ \hline
\begin{tabular}[c]{@{}c@{}}{BIM-a}\\ {and}\\ {BIM-b}\end{tabular} & \begin{tabular}[c]{@{}c@{}}{Decision}\\ {boundary}\\ {adaptability}\end{tabular} & \begin{tabular}[c]{@{}c@{}}{Not necessarily}\\ {optimal in}\\ {runtime}\end{tabular} & {Non-targeted}                                                       & $l_{\infty}$                                                & {Iterative}                                                     \\ \hline
{JSMA}                                                        & \begin{tabular}[c]{@{}c@{}} {Jacobian}\\ {matrix-based}\end{tabular}            & \begin{tabular}[c]{@{}c@{}}{Costly in}\\ {saliency map}\end{tabular}               & {Targeted}                                                           & $l_{0}$                                                     & {Iterative}                                                     \\ \hline
{CWA}                                                         & \begin{tabular}[c]{@{}c@{}} {Strong and}\\ {natural attack}\end{tabular}        & \begin{tabular}[c]{@{}c@{}} {Complex}\\ {transformations}\end{tabular}              & {Targeted}                                                           & $l_{0}$, $l_{2}$, $l_{\infty}$                              & {Iterative}                                                     \\ \hline
\end{tabular}
\label{table:propertiesCompbench}
\end{table*}

\section{2D Representations for Audio Signals}
\label{audio_rep_section}
Representing audio signals using time-frequency plots is a standard operation in audio and speech processing representing such signals in a compact and informative way. Fourier and wavelet transforms are the most commonly used approaches for mapping an audio signal into frequency-magnitude representations.

Two popular representations derived from Fourier transform are STFT and MFCC, where the latter is the condensed version of the first with lower dimensionality. However, they both have straightforward algorithms. In a nutshell, they firstly divide the input 1D signal into small frames and then apply the Fourier transform to obtain frequency-magnitude coefficients \cite{hasan2004speaker}. Finally, they run some postprocessing operations (e.g., non-linear transformation) to filter out and organize the achieved features. Detailed information about these two types of spectrograms is available in \cite{brigham1988fast,benesty2008springer}.

In terms of functionality, DWT is very similar to the two representations mentioned above. However, it convolves the input signal with a mother function to obtain the spectral features \cite{pathak2009wavelet}. Moreover, this function is often complex and symmetrical to extract more informative features \cite{daubechies1993ten}.

Although discussion on the comparison of these transformations is out of the scope of this paper, we mention one of their critical differences. Whereas the Fourier-based representations, wavelet transform not only extracts the local feature of the signal, but also it can provide the exact location of the features \cite{pathak2009wavelet}. However, in terms of computational complexity, DWT is relatively costlier than MFCC and STFT \cite{boussaa2016comparison}.

In the next section, we explain our experiments on three benchmarking sound datasets. We firstly generate separate spectrogram sets with the three representations mentioned above using different configurations. Second, we train a ResNet on these datasets and run adversarial attack algorithms against them. Finally, we measure both the fooling rate and the cost of attacks. We demonstrate that for different spectrogram configurations, these metrics are meaningfully different. 

\section{Experiments}
\label{sec:exp}
We use three environmental sound datasets in all our experiments: UrbanSound8k \cite{Salamon:UrbanSound:ACMMM:14}, ESC-50 \cite{piczak2015esc}, and ESC-10 \cite{piczak2015esc}. The first dataset includes 8732 four-second length audio samples distributed in 10 classes: engine idling, car horn, children playing, drilling, air conditioner, jackhammer, dog bark, siren, gunshot, and street music. ESC-50 is a comprehensive dataset with 50 different classes and overall 2000 five-second audio recordings of natural acoustic sounds. A subset of this dataset is ESC-10 which has been released with ten classes and 400 recordings.

For increasing both the quality and the number of samples of these datasets, we apply a pitch-shifting augmentation approach with scales $0.75$, $0.9$, $1.15$, and $1.5$ as proposed in \cite{esmaeilpour2019unsupervised}, which positively affect classification accuracy. This data augmentation operation generates four extra audio samples for every original audio sample, and eventually, it increases the size of the original dataset by the factor of four. We discuss the usefulness of this 1D data augmentation approach in Section~\ref{discuss:sec}. In the following subsection, we explain the details of generating 2D representations for audio signals. Toward this aim, we use the open-source Librosa signal processing python library \cite{mcfee2015librosa} and our upgraded version of the wavelet toolbox \cite{waveletSoundExplorer}.

\subsection{Generating Spectrograms}
For every dataset including augmented signals, we separately generate independent sets of 2D representations, namely MFCC, STFT, and DWT. We aim to investigate which audio representation yields a better trade-off between recognition accuracy and robustness for a victim model against various strong adversarial attacks.

\subsubsection{MFCC Production Settings}
\label{sec:advAttForMFCCPB}
There are four major settings in generating MFCC spectrogram using Librosa. The default value for sampling rate is 22.05 kHz. Since there is no optimal approach for determining the best sampling rate, we generate the most informative spectrogram. We run extensive experiments using sampling rates from 8 to 24 kHz. The second tunable hyperparameter is the number of MFCCs ($N_{\text{MFCC}}$), which we examine different values for it: 13, 20, and 40 per frame with a hop length of 1024. Normalization of discrete cosine transform (type 2 or 3) using orthonormal DCT basis for MFCC production is the third setting. By default, this hyperparameter is set to true in almost all the libraries, including Librosa. However, we measure the performance of the front-end classifier trained to MFCC spectrograms without normalization. The last argument is about the number of cepstral filtering ($CF$) \cite{juang1987use} to be applied on MFCC features. The sinusoidal $CF$ reduces involvement of higher-order coefficients and improve recognition performance \cite{paliwal1999decorrelated} (see Eq.~\ref{lifter_cf}).
\begin{equation}
 \bold{M}\left [ n,: \right ] \leftarrow \bold{M}\left [ n,: \right ] \times \left ( 1+\sin \left( \frac{\pi(n+1)}{CF} \right) \right ) \frac{CF}{2}
 \label{lifter_cf}
\end{equation}
\noindent where $\bold{M}$ stands for MFCC array with size $[n,:]$. We investigate the effect of $CF$ on the overall performance of classification models.

\subsubsection{STFT Production Settings}
\label{sec:advAttForSTFTPB}
For producing STFT representations, we use default configurations for general hyperparameters as outlined in the Librosa manual. We use 2048, 1024, and 512 with associated sampling rates for assigning the length of the windowed signal. We also use variable window sizes: 2048 (default value), 1024, and 512 (very small window) associated with a default hop size of 512. We investigate the potential effects of these configurations for the resiliency of the victim models against adversarial attacks.

\subsubsection{DWT Production Settings}
\label{sec:advAttForDWTPB}
For generating DWT representations, we modified the sound explorer software \cite{waveletSoundExplorer} to support Haar and Mexican Hat wavelet mother functions in addition to complex Morlet. Sampling frequency for DWT spectrograms has been set up to 8 kHz and 16 kHz with a constant frame length of 50 ms. Moreover, by convention, the overlapping threshold is set to 50\%. Our experiments measure the impacts of these DWT configurations visualized in logarithmic scale (for higher resolution) on both recognition accuracy and robustness against adversarial attacks. 

In the following subsection, we discuss possible choices for the classification models to be separately trained on the spectrogram representations and setups mentioned above. Finally, we select our final front-end classifier from a diverse domain of traditional handcrafted-based feature learning algorithms to state-of-the-art DL architectures.

\subsection{Classification Model}
\label{classification_model_Sec}
For the choice of classification algorithms, we initially included both conventional classifiers such as linear and Gaussian SVM \cite{esmaeilpour2019robust}, random forest \cite{esmaeilpour2019unsupervised}, and some deep learning architectures. Specifically, we selected pre-trained GoogLeNet (because of its inception mechanism), AlexNet (for taking advantage of its fully convolutional configuration), and ResNet (utilizing a mixture of residual and convolutional blocks) \cite{he2016deep} models tuned for our three benchmarking datasets. We preserved the architectures of these ConvNets except for the first layer and the last layer for mapping logits into class labels (softmax layer). Since spectrograms may have different dimensions according to their length and transformation schemes, we bilinearly interpolate them to fit 128$\times$128 for all the ConvNets.

Performance comparison of the SVMs, GoogLeNet, and AlexNet mentioned above against a few adversarial attacks have already been studied mainly for DWT representations of environmental sound datasets in \cite{esmaeilpour2019robust}. However, their experiments have been conducted on standard spectrograms without validating the potential impacts of different settings in producing different representations. In this paper, we carry out extensive experiments using: (i) three common 2D representations for audio signals, namely MFCC (represented in 2D matrix format, not the common vector visualization), STFT, and DWT; (ii) more and stronger targeted and non-targeted algorithms for adversarial attacks; (iii) fair comparison on fooling rates of victim models taking their cost of attacks averaged over the allocated budgets into account. 

We primarily select a ConvNet as our front-end classifier for the sake of simplicity and interpretability of results. We present concise results for other classification models in Section~\ref{discuss:sec}. We selected ResNet architectures for such an aim because such a ConvNet is currently the best-performing classifier for several tasks \cite{hershey2017cnn}. Our implementations corroborate that, on average, these ConvNet architectures outperform all the algorithms mentioned above (both SVMs and other DL approaches) trained on spectrograms. Among the possible architectures for ResNet (ResNet-18, ResNet-34, and ResNet-56), we selected ResNet-18 according to its highest recognition performance and relatively low number of parameters compared to others. Recalling that, we investigate the potential effects of spectrogram configurations on the classifier, which has a very competitive recognition accuracy compared to others and requires fewer training parameters. Thus, we specifically focus on the ResNet-18 network, and all our investigations will consider this victim architecture.

For every configuration to produce the 2D representations, we generate an individual set of spectrograms and train an independent ResNet-18 classifier on each dataset. We use a 5-fold cross-validation setup on 70\% of the overall dataset volume (training plus development). We implemented the early stopping technique in training to avoid overtraining and finally reported mean recognition accuracy on the test sets (30\% remaining).

\subsection{Adversarial Attacks}
\label{adv:setup}
In this section, we provide details for attacking the models trained on audio representations. We examine their robustness against six strong adversarial attacks by reporting obtained average model robustness ratio using two metrics of the area under the region of convergence (ROC \cite{gradshteyn2014table}) curve (AUC) \cite{cortes2005confidence} and the total number of gradient computations.

Model robustness refers to the average recognition accuracy of the victim classifier evaluated on the adversarial examples (spectrograms in our case) \cite{ma2018characterizing}. In other words, it measures the ratio of correctly classified adversarial spectrograms over the total number of crafted examples using the AUC metric. It is worth mentioning that there is an inverse relationship between the model robustness and attack fooling rate. More specifically, the latter measures the ratio of misclassified adversarial spectrograms over the total number of crafted examples (see \cite{ma2018characterizing} for more details).

To the best of our knowledge, all the adversarial attack algorithms are optimization-based procedures toward achieving the minimum possible perturbation. These procedures should generate spectrograms very similar to the ground-truth using a specific similarity metric. This metric is often one of the statistical norms such as $l_{0}$, $l_{2}$, $l_{\infty}$, etc. \cite{goodfellow2014explaining}. Thus, attack algorithms should minimize over the designated similarity metric in an iterative pipeline. The total number of times (in each batch) which this pipeline should be executed until achieving a valid (in terms of being far enough from the decision boundary of the victim model \cite{papernot2018characterizing}) adversarial spectrogram is called gradient computation or callback to the ground-truth. This process imposes considerable computational overhead to the entire attack optimization procedure and limits the adversary’s strength in runtime. Therefore, increasing the number of required gradient computations is a potential way to decrease the fooling rate of the victim model and potentially resist attacks.

Thus far, it has been demonstrated that the fooling rate of a classifier is dependent on the properties of the attack algorithm, the allocated budget in runtime, and the characteristics of the victim model \cite{papernot2018characterizing}. For instance, some attack algorithms (e.g., CWA) can get closer to the decision boundary of the victim classifier and consequently find a smaller adversarial perturbation. This results in more effectively attacking the recognition model and increases the fooling rate. Furthermore, since changing the settings of the spectrograms modifies the decision boundary of the audio classifiers, it will most likely affect the fooling rate of the victim model.

\subsubsection{Settings for Attack Algorithms}
In FGSM and BIMs attacks, possible ranges for $\delta$ have been defined from $0.001$ to any possible supremum under different confidence intervals ($\geq 65\%$). For the implementation of the DeepFool attack, we use the open-source Foolbox package \cite{rauber2017foolbox} with iterations from 100 to 1000 (10 different scales with a step of 100). In the implementation of the JSMA attack, the number of iterations has been set to $(m_{i} \gamma)/n_{i}$ where $m_{i}$ and $n_{i}$ denote the total number of pixels and scaling factor within $[0, 200]$ (with displacement a of 40), respectively. Also $\gamma$ is the maximum allowed distortion (ideally $< 1.5/255$) within the maximum number of iterations. Budget allocated to CWA is within $\left \{1, 3, 7, 9 \right \}$ for search steps in $c$ within $\left \{25, 100, 1\mathrm{k}, 2\mathrm{k}, 5\mathrm{k} \right \}$ iterations in each search step using early stopping. For targeted attacks (i.e., FGSM, JSMA, and CWA) we randomly select targeted wrong labels for running adversarial optimization algorithms.

We executed these attack algorithms on two NVIDIA GTX-1080-Ti with $4 \times 11$ GB of memory except for the DeepFool attack, which was executed on 64-bit Intel Core-i7-7700 (3.6 GHz) CPU with 64 GB memory. For attacks on the smallest dataset (ESC-10), we used batches of 200 samples. For larger datasets (ESC-50 and UrbanSound8k), we used 25 batches of 100 samples.

\subsubsection{Adversarial Attacks for MFCC Representations}
We firstly investigate the potential effect of different sampling rates in MFCC production on the performance of the trained models. To this end, sampling rates have been selected from reasonably low (8 kHz) to moderately high (24 kHz) ranges, including the default frequency value (22.05 kHz) defined in Librosa. Therefore, we trained four ResNet-18 models per dataset associated with four sampling rates. The results summarized in Table~\ref{mfcc_sr_effect} show that the recognition performance of the classifiers is, to some extent, dependent on the sampling rates. For example, for ESC-10 and UrbanSound8k datasets, the sampling rate of 8 kHz improves recognition accuracy, while 16 kHz works better for ESC-50. These results might imply that a high sampling rate filters out low-frequency components and negatively affects the learning of discriminative features from the spectrograms. 

\begin{table}
\centering
\scriptsize
\setlength{\tabcolsep}{0.5em}
\caption{Performance comparison of models trained on MFCC representations with different sampling rates averaged over experiments and budgets. Relatively better performances are in boldface.}
\begin{tabular}{|c|c|c||c|c|c|c|c|c|}
\hline
\multirow{2}{*}{Dataset} & Sampling & Recog. & \multicolumn{6}{c|}{AUC Score for Fooling Rate, Number of Gradients for Adversarial Attacks} \\ \cline{4-9} 
 & Rate (kHz) & Acc. (\%) & FGSM & DeepFool & BIM-a & BIM-b & JSMA & CWA \\ \hline \hline
\multirow{4}{*}{ESC-10} & 8 & \textbf{73.23} & \textbf{0.9822}, 1 & 0.9473, 074 & \textbf{0.9710}, 065 & \textbf{0.9801}, 110 & \textbf{0.9308}, 096 & \textbf{0.9912}, 1346 \\ \cline{2-9} 
 & 16 & 72.15 & 0.9456, 1 & \textbf{0.9607}, 046 & 0.9334, 059 & 0.9375, 197 & 0.9144, 151 & 0.9616, 1435 \\ \cline{2-9} 
 & 22.05 & 72.06 & 0.9467, 1 & 0.9518, 129 & 0.9309, 088 & 0.9379, 186 & 0.9145, 213 & 0.9405, 1471 \\ \cline{2-9} 
 & 24 & 70.13 & 0.9471, 1 & 0.9341, 078 & 0.9298, 115 & 0.9327, 171 & 0.9233, 091 & 0.9302, 1149 \\ \hline \hline
\multirow{4}{*}{ESC-50} & 8 & 69.89 & 0.9517, 1 & 0.9023, 061 & 0.9612, 084 & 0.9703, 193 & 0.9288, 118 & 0.9598, 2418 \\ \cline{2-9} 
 & 16 & \textbf{70.21} & \textbf{0.9849}, 1 & \textbf{0.9912}, 248 & \textbf{0.9871}, 209 & \textbf{0.9903}, 160 & \textbf{0.9508}, 251 & 0.9672, 2639 \\ \cline{2-9} 
 & 22.05 & 69.97 & 0.9534, 1  & 0.9386, 331 & 0.9430, 423 & 0.9581, 288 & 0.9233, 219 & 0.9434, 2318 \\ \cline{2-9} 
 & 24 & 67.25 & 0.9433, 1 & 0.9214, 208 & 0.9307, 159 & 0.9415, 216 & 0.9187, 417 & \textbf{0.9652}, 2744 \\ \hline \hline
\multirow{4}{*}{US8k} & 8 & \textbf{71.25} & \textbf{0.9905}, 1 & \textbf{0.9895}, 326 & 0.9411, 317 & \textbf{0.9950}, 223 & \textbf{0.9623}, 398 & \textbf{0.9708}, 2791 \\ \cline{2-9} 
 & 16 & 70.81 & 0.9508, 1 & 0.9215, 631 & 0.9346, 519 & 0.9389, 817 & 0.9447, 442 & 0.9449, 3805 \\ \cline{2-9} 
 & 22.05 & 69.57 & 0.9457, 1 & 0.9151, 269 & \textbf{0.9449}, 184 & 0.9256, 513 & 0.9370, 416 & 0.9456, 3015 \\ \cline{2-9} 
 & 24 & 69.33 & 0.9440, 1 & 0.9221, 318 & 0.9236, 299 & 0.9120, 862 & 0.9242, 343 & 0.9371, 2816 \\ \hline
\end{tabular}
\label{mfcc_sr_effect}
\end{table}

We attack these models using those six adversarial algorithms mentioned above and measure their fooling rates averaged over different budgets as explained in Section \ref{adv:setup}. From the results shown in Table~\ref{mfcc_sr_effect}, we notice an inverse relationship between recognition accuracy and robustness of these models, on average. For instance, ResNet-18 trained on MFCC spectrograms of the ESC-10 dataset sampled at 8 kHz reaches the highest recognition accuracy. Still, this model is less robust against five out of six adversarial attacks, averaged over the allocated budgets. We present two hypotheses on this issue. Firstly, adversarial attacks are essentially optimization-based problems, and their final results depend on the hyperparameters defined by the adversary. Confidence intervals, number of callbacks to the original spectrogram, number of iterations in optimization formulation, line search for the optimal coefficient are among those, to name a few. Hence, the fooling rate of a victim model is dependent on tuning these hyperparameters. Our second hypothesis is on the statistical perspective of training a neural network. A model with higher recognition accuracy has probably learned a better decision boundary via maximizing the intra-class similarity and inter-class dissimilarity. Hence, attacking this model provides a broader search space for the adversary to find pinholes of the model, especially when the decision boundaries among classes lie in the vicinity of each other. Table~\ref{mfcc_sr_effect} also compares the average number of gradients for batch execution required by every attack algorithm. Regarding the statistics of this table, CWA is the costliest adversarial attack for spectrograms with different sampling rates.

\begin{figure}[htpb!]
  \centering
  \includegraphics[width=0.75\textwidth]{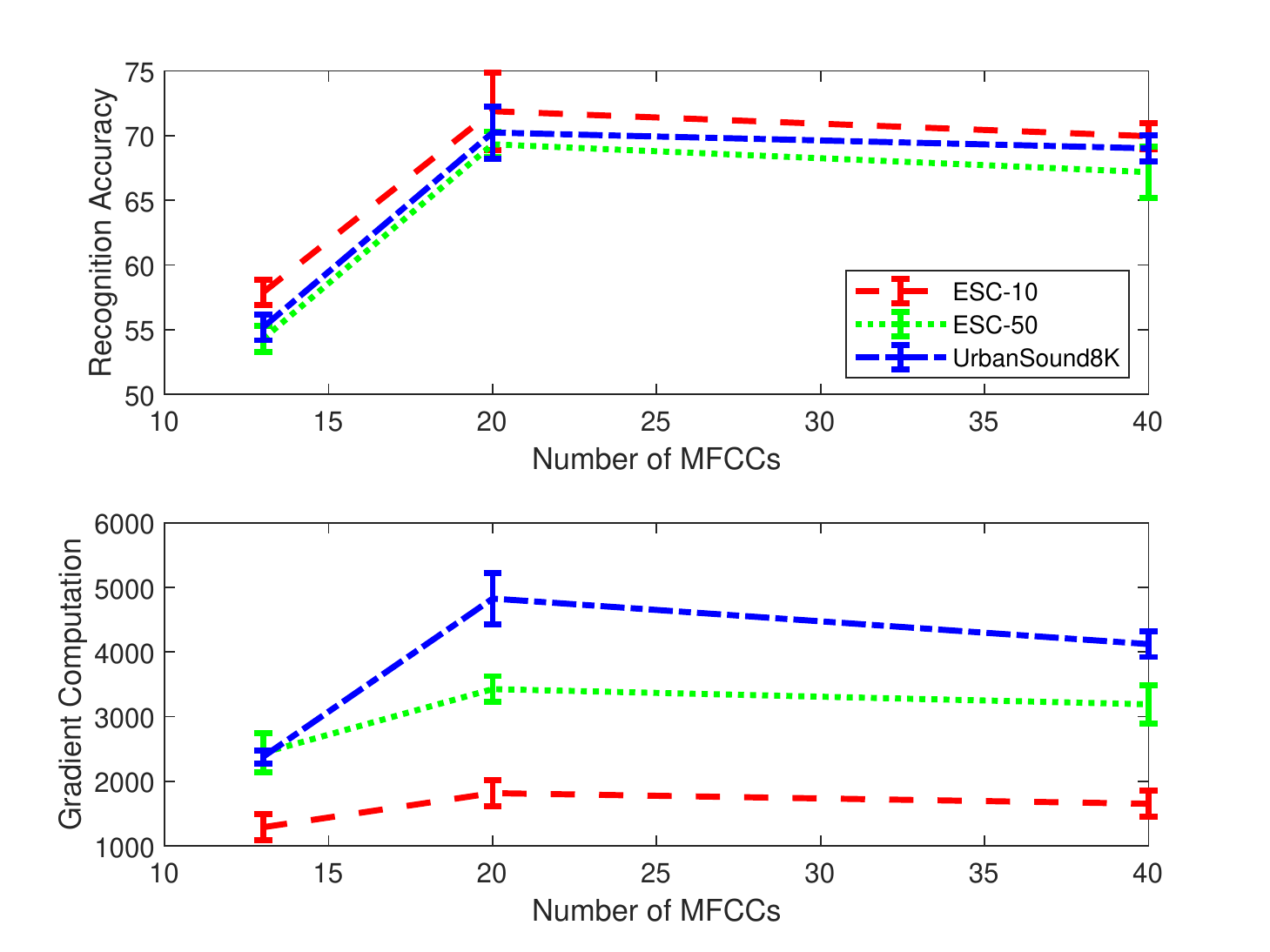}
  \caption{Effect of $N_{\mathrm{MFCC}}$ on the recognition accuracy and the average cost of the attack (number of batch gradient computation) over six adversarial algorithms for ResNet-18 models.}
  \label{mel-attack-pack}
\end{figure}

The default value for the number of MFCCs ($N_{\text{MFCC}}$) is 20 as defined in Librosa. However, we encompass values from a minimum number of 13 to a maximum of 40 in generating MFCC spectrograms; although increasing $N_{\text{MFCC}}$$>$20 introduces redundancy in frequency coefficient representation. Our experimental results corroborate the negative effect of a low $N_{\text{MFCC}}$ in the performance of the classifiers. More specifically, recognition performance of the trained models on spectrograms with $N_{\text{MFCC}}=$ 13 is 14\% less than models trained on spectrograms with $N_{\text{MFCC}}$$\geq$20, on average. Our experimental results on attacking victim models trained on spectrograms with low $N_{\text{MFCC}}$ unveil their extreme vulnerabilities. However, in terms of the attack cost, these models need fewer callbacks for gradient computations for yielding AUC$>$90\%  (see Figure~\ref{mel-attack-pack}). That could be due to the nature of the adversarial attacks, which are formulated as optimization problems, regardless of the performance of the victim models.

\begin{figure}[htpb!]
  \centering
  \includegraphics[width=0.75\textwidth]{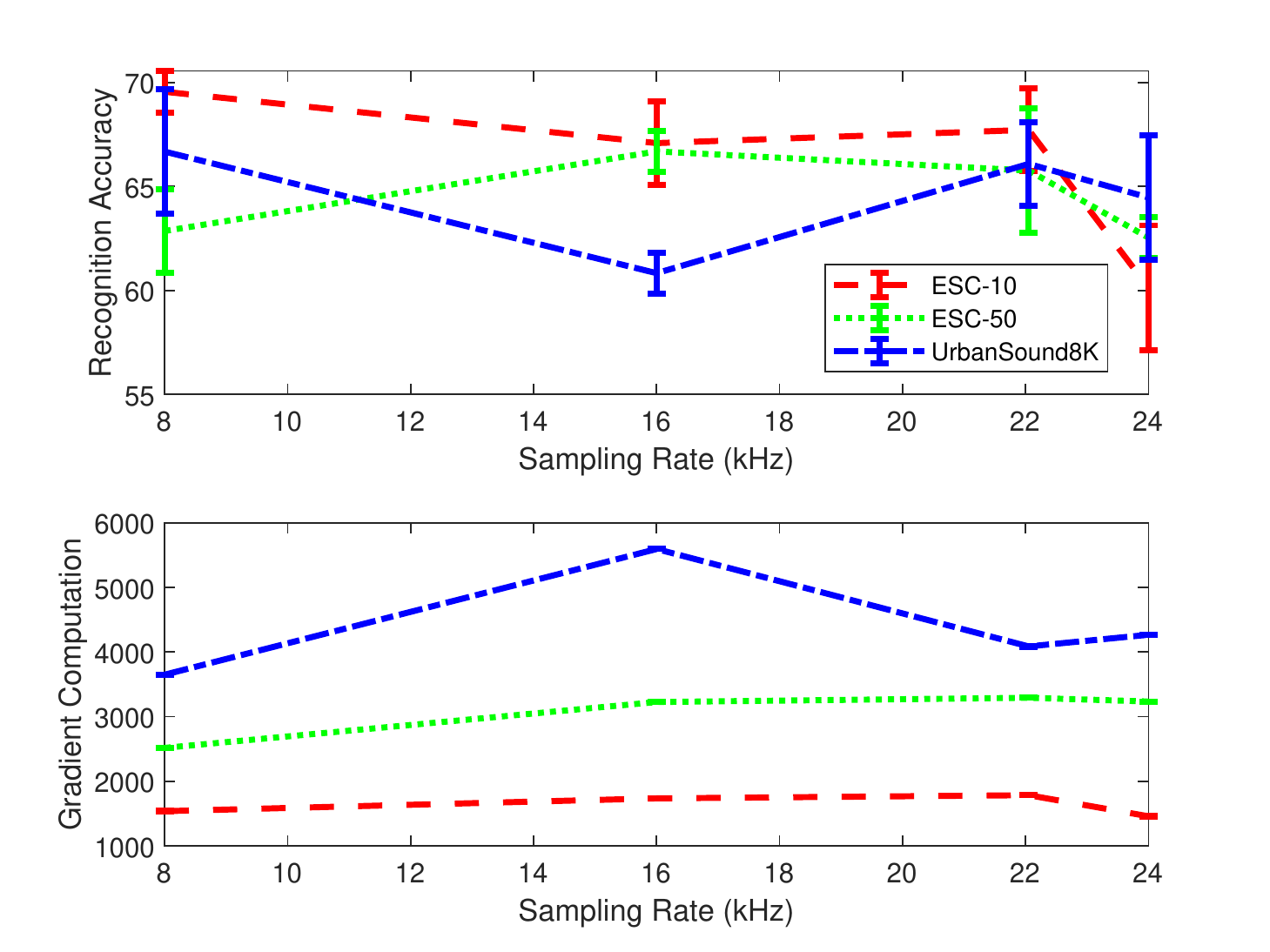}
  \caption{{Comparison of recognition accuracy and gradient computation at each sampling rate for normalized spectrograms. Herein, gradients computations denote the attack cost for reaching $\mathrm{AUC} > 0.9$ (this threshold indicates a reasonably high vulnerability \cite{papernot2018characterizing,sallo2021adversarially}) over six adversarial algorithms for ResNet-18 models.} }
  \label{mel-attack-pack2}
\end{figure}

Using the orthonormal discrete cosine transform basis function is a standard approach in crafting MFCC spectrograms. Our experiments produced two separate subsets of spectrograms with and without normalization to measure its potential effect on recognition accuracy and the fooling rate. Figure~\ref{mel-attack-pack2} tracks the relation among sampling rate, recognition accuracy, and attack cost for normalized spectrograms. Disabling this normalization scheme causes a drop of 7\% in the recognition accuracy and 8.5\% in the attack cost, on average.

\begin{figure}[htpb!]
  \centering
  \includegraphics[width=0.75\textwidth]{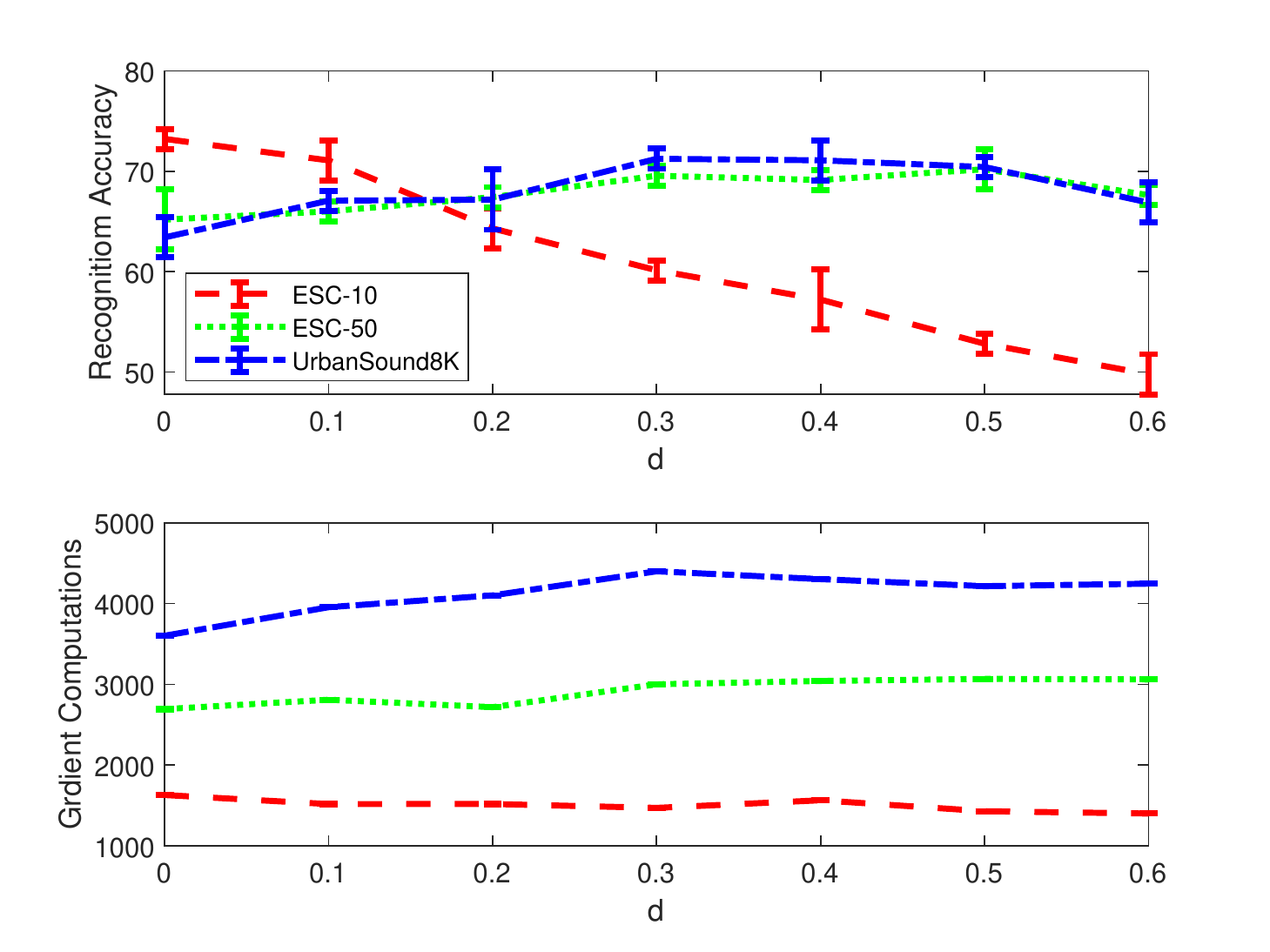}
  \caption{Effect of Cepstral filtering on the recognition accuracy and the average attack cost for reaching $AUC>0.9$ over six adversarial algorithms for ResNet-18 models.}
  \label{mel-attack-pack3}
\end{figure}

For the choice of the cepstral filtering, we covered values in the range $\begin{bmatrix} 0, \left (d \times N_{\text{MFCC}} \right ) \end{bmatrix}$, where the maximum $d$ is 2.5 with a hop size of 0.5 in the production of spectrograms. Values above the supremum of this interval generate higher-order coefficients in linear-like weighting distributions, which considerably reduce recognition accuracy on average to about 48\%. Figure~\ref{mel-attack-pack3} shows the effect of the $d$ parameter on both the recognition accuracy and gradient computations.

\subsubsection{Adversarial Attacks for STFT Representations}
\label{sec:advAttForSTFTPP}
There is a significant similarity in producing MFCC and STFT spectrograms, mainly in terms of transformation and frequency modulation. Therefore, we omit experimental results relevant to measuring the impacts of sampling rates on the robustness of victim classifiers. Nevertheless, fooling rates of ResNet-18 models on STFT representations are similar to MFCC representations. Such rates support the inverse relationship between the recognition accuracy and the robustness against attacks mentioned above.

\begin{figure}[htpb!]
  \centering
  \includegraphics[width=0.75\textwidth]{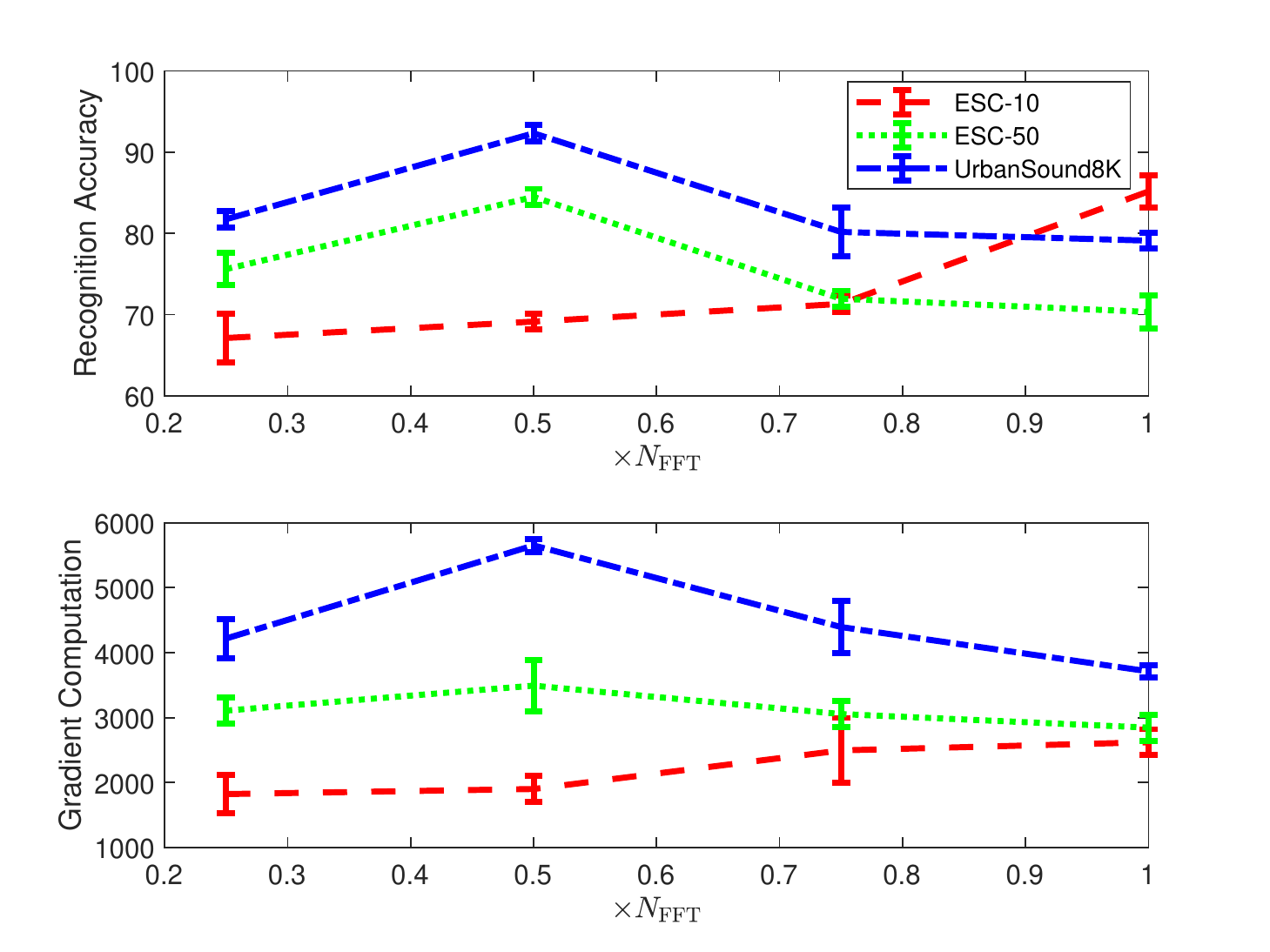}
  \caption{Effect of scales for $N_{\mathrm{FFT}}$ on the recognition accuracy and on the average cost of the attack for reaching $AUC>0.9$ over six adversarial algorithms for ResNet-18 models.}
  \label{mel-attack-pack4}
\end{figure}

Table~\ref{stft_fft_effect} summarizes adversarial experiments conducted on STFT representations with the same aforementioned setup described in Section~\ref{adv:setup}. This table illustrates the impact of the number of FFTs ($N_{\text{FFT}}$) both on the recognition accuracy and on the robustness of victim models against adversarial attacks averaged over all the different adversarial setups. For ESC-10 and ESC-50 datasets, $N_{\text{FFT}}$$=$1024 results in learning better decision boundaries for the classifiers, although it increases fooling rates of the victim models. In the production of STFT spectrograms, each frame of a given audio signal is spanned by a window that covers the frame. The maximum length of this window can be equivalent to the number of $N_{\text{FFT}}$. Since small window lengths improve the temporal resolution of the final STFT representation, we evaluate the performance of the models on small window lengths in the range $\begin{bmatrix}\begin{pmatrix}0.25$$\times$$N_{\text{FFT}}\end{pmatrix}, N_{\text{FFT}}\end{bmatrix}$ with hop size of $N_{\text{FFT}}/4$. As shown in Figure~\ref{mel-attack-pack4}, the evaluation on ESC-50 and UrbanSound8k datasets uncovers that models trained on STFT representations with window length of $0.5$$\times$$N_{\text{FFT}}$ outperform others. On the ESC-10 dataset, a window length of $N_{\text{FFT}}$ resulted in better performance in terms of recognition accuracy.

\begin{table}
\centering
\scriptsize
\setlength{\tabcolsep}{0.5em}
\caption{Performance comparison of models trained on STFT representations with different $N_{\text{FFT}}$ averaged over experiments and budgets. Relatively better performances are in boldface.}
\begin{tabular}{|c|c|c||c|c|c|c|c|c|}
\hline
\multirow{2}{*}{Dataset} & Number & Recog. & \multicolumn{6}{c|}{AUC Score for Fooling Rate, Number of Gradients for Adversarial Attacks} \\ \cline{4-9} 
 & of FFTs & Acc. (\%) & FGSM & DeepFool & BIM-a & BIM-b & JSMA & CWA \\ \hline \hline
\multirow{3}{*}{ESC-10} & 512 & 82.41 & 0.9768, 1 & 0.9430, 089 & 0.9576, 109 & 0.9717, 134 & 0.9662, 141 & 0.9846, 1415 \\ \cline{2-9} 
 & 1~024 & \textbf{85.17} & \textbf{0.9823}, 1 & \textbf{0.9701}, 129 & \textbf{0.9715}, 091 & \textbf{0.9792}, 183 & 0.9531, 209 & \textbf{0.9905}, 2008 \\ \cline{2-9} 
 & 2~048 & 80.56 & 0.9651, 1 & 0.9544, 092 & 0.9407, 163 & 0.9529, 279 & 0.9588, 341 & 0.8731, 1730 \\ \hline \hline
\multirow{3}{*}{ESC-50} & 512 & 82.44 & 0.9786, 1 & 0.9542, 082 & 0.9583, 109 & 0.9665, 244 & 0.9614, 128 & 0.9618, 1995 \\ \cline{2-9} 
 & 1~024 & \textbf{84.49} & \textbf{0.9881}, 1 & 0.9512, 331 & \textbf{0.9871}, 267 & \textbf{0.9798}, 179 & 0.9702, 361 & \textbf{0.9896}, 2353 \\ \cline{2-9} 
 & 2~048 & 83.12 & 0.9567, 1 & \textbf{0.9631}, 145 & 0.9765, 211 & 0.9606, 567 & \textbf{0.9738}, 399 & 0.9729, 2412 \\ \hline \hline
\multirow{3}{*}{US8k} & 512 & 90.58 & 0.9761, 1 & 0.9414, 583 & \textbf{0.9513}, 442 & 0.9682, 421 & 0.9402, 345 & 0.9539, 2569 \\ \cline{2-9} 
 & 1~024 & 91.74 & 0.9827, 1 & 0.9752, 322 & 0.9340, 471 & \textbf{0.9687}, 719 & 0.9515, 502 & 0.9654, 3271 \\ \cline{2-9} 
 & 2~048 & \textbf{92.23} & \textbf{0.9895}, 1 & \textbf{0.9764}, 643 & 0.9407, 602 & 0.9630, 408 & \textbf{0.9623}, 655 & \textbf{0.9673}, 3342 \\ \hline
\end{tabular}
\label{stft_fft_effect}
\end{table}

\begin{table}[t]
\centering
\scriptsize
\setlength{\tabcolsep}{0.5em}
\caption{Performance comparison of models trained on DWT representations with different sampling rates averaged over different budgets. Relatively better performances are in boldface.}
\begin{tabular}{|c|c|c||c|c|c|c|c|c|}
\hline
\multirow{2}{*}{Dataset} & \multirow{2}{*}{\begin{tabular}[c]{@{}c@{}}Sampling\\ Rate (kHz)\end{tabular}} & \multirow{2}{*}{\begin{tabular}[c]{@{}c@{}}Recog.\\ Acc. (\%)\end{tabular}} & \multicolumn{6}{c|}{AUC Score for Fooling Rate, Number of Gradients for Adversarial Attacks} \\ \cline{4-9} 
 & & & FGSM & DeepFool & BIM-a & BIM-b & JSMA & CWA \\ \hline \hline
\multirow{2}{*}{ESC-10} & 8 & \textbf{85.67} & \textbf{0.9456}, 1 & \textbf{0.9310}, 429 & 0.9307, 612 & \textbf{0.9411}, 744 & \textbf{0.9324}, 781 & \textbf{0.9483}, 4205 \\ \cline{2-9} 
 & 16 & 82.04 & 0.9068, 1 & 0.9192, 672 & \textbf{0.9437}, 490 & 0.9347, 513 & 0.9018, 801 & 0.9216, 4439 \\ \hline \hline
\multirow{2}{*}{ESC-50} & 8 & 80.34 & \textbf{0.9462}, 1 & \textbf{0.9335}, 367 & 0.9161, 452 & 0.9314, 809 & 0.9168, 298 & 0.9233, 3981 \\ \cline{2-9} 
 & 16 & \textbf{85.97} & 0.9376, 1 & 0.9256, 409 & \textbf{0.9314}, 628 & \textbf{0.9419}, 701 & \textbf{0.9173}, 561 & \textbf{0.9236}, 4575 \\ \hline \hline
\multirow{2}{*}{US8k} & 8 & \textbf{94.70} & \textbf{0.9401}, 1 & \textbf{0.9279}, 761 & \textbf{0.9315}, 841 & \textbf{0.9511}, 738 & \textbf{0.9207}, 691 & 0.9320, 4684 \\ \cline{2-9} 
 & 16 & 91.83 & 0.9321, 1 & 0.9274, 533 & 0.9125, 719 & 0.9408, 941 & 0.9139, 774 & \textbf{0.9430}, 4879 \\ \hline
\end{tabular}
\label{dwt_effect}
\end{table}
Comparing the recognition accuracy of Tables~\ref{mfcc_sr_effect} and~\ref{stft_fft_effect} shows that STFT provides better discriminative features for the ResNet-18 classifier since such a model achieved lower recognition accuracy on MFCC representations. Additionally, while the $AUC$ scores across the six attacks are not so different, ranging from 0.93 to 0.99, the number of gradients required for models trained on STFT spectrograms is considerably higher than MFCC spectrograms. In summary, STFT spectrograms provide better accuracy and are a little hard to attack, even if they can be fooled with high success by all six adversarial attacks.

\subsubsection{Adversarial Attacks for DWT representations}
There is no algorithmic approach for obtaining the optimal mother function to generate DWT spectrograms. Therefore, from simple Haar to complex Morlet, we have employed several functions to investigate the potential impacts on recognition accuracy and the adversarial robustness of the victim models. In addition, we exploited an analytical approach, recasting multiple experiments. Table~\ref{mother_func_comp} shows that although the complex Morlet mother function outperforms other mother functions in terms of recognition accuracy. However, it shows more vulnerability against adversarial examples, averaged over six attack algorithms with different budgets.
\begin{table}[htpb!]
\centering
\scriptsize
\setlength{\tabcolsep}{0.5em}
\caption{Comparison of mother functions on the performance of the models. Outperforming values are shown in bold face. Results are averaged over a comprehensive set of mother functions with different time decay (around 0.01) and regular grid parameters \cite{teolis1998computational}.}
\begin{tabular}{|c||c|c|c|}
\hline
Dataset & \begin{tabular}[c]{@{}c@{}}Mother\\ Function\end{tabular} & \begin{tabular}[c]{@{}c@{}}Average Recognition\\ Accuracy (\%)\end{tabular} & \begin{tabular}[c]{@{}c@{}}Average\\ AUC Score\end{tabular} \\ \hline \hline
\multirow{3}{*}{ESC-10} & Haar & 82.14 & 95.14 \\ \cline{2-4} 
 & Mexican Hat & 84.51 & 94.19 \\ \cline{2-4} 
 & Complex Morlet & \textbf{85.67} & \textbf{95.61} \\ \hline \hline
\multirow{3}{*}{ESC-50} & Haar & 83.08 & 92.16 \\ \cline{2-4} 
 & Mexican Hat & 84.33 & 93.40 \\ \cline{2-4} 
 & Complex Morlet & \textbf{85.97} & \textbf{95.38} \\ \hline \hline
\multirow{3}{*}{UrbanSound8k} & Haar & 91.22 & 96.16 \\ \cline{2-4} 
 & Mexican Hat & 93.48 & 95.63 \\ \cline{2-4} 
 & Complex Morlet & \textbf{95.17} & \textbf{96.09} \\ \hline
\end{tabular}
\label{mother_func_comp}
\end{table}

Table~\ref{dwt_effect} compares the recognition accuracy of models trained on DWT representations with complex Morlet mother function. We have evaluated these models on DWT spectrograms with sampling rates of 8 kHz and 16 kHz. For ESC-50, a sampling rate of 8 kHz shows better performance for the classifiers, comparing their recognition accuracy. There are three findings in these tables. Firstly, averaged over all the allocated budgets for the attacks, models trained on DWT representations demonstrate slightly higher robustness against adversarial attacks than MFCC and STFT spectrograms. Secondly, the highest recognition accuracy has been achieved for classifiers trained on DWT representations. Thirdly, the trade-off between recognition accuracy and adversarial robustness of the victim models are noticeable for different sampling rates. Moreover, the cost of the attack (number of gradient computations) for models trained on DWT is considerably higher than the other two representations. 

\begin{figure}[htpb!]
  \centering
  \includegraphics[width=0.75\textwidth]{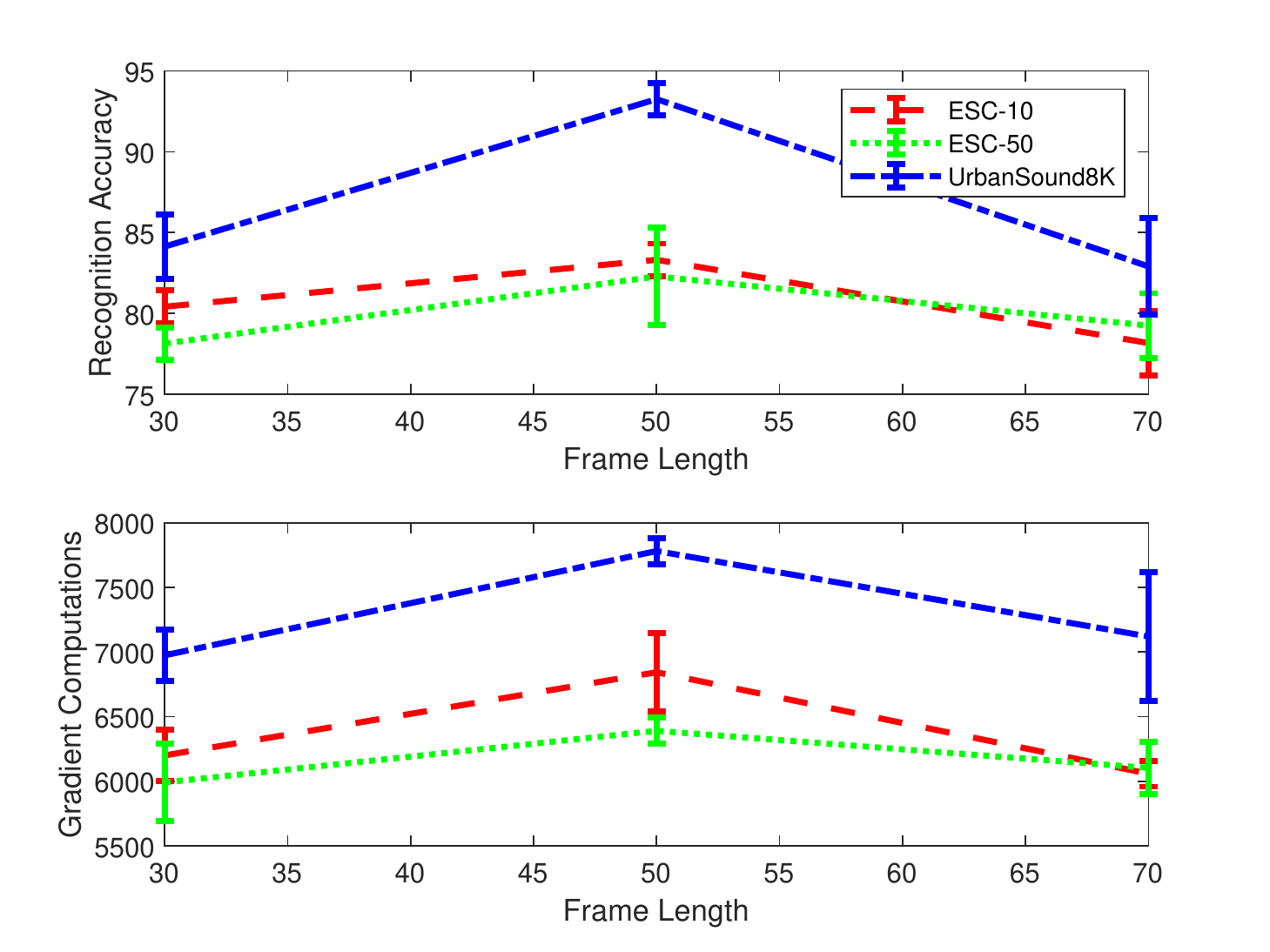}
  \caption{The effect of DWT frame length on the recognition accuracy and on the average cost of the attack for yielding $AUC>0.9$ over six adversarial algorithms for ResNet-18 models.}
  \label{mel-attack-pack5}
\end{figure}

\begin{sidewaysfigure}
 \footnotesize 
\setlength{\tabcolsep}{2pt}
 \begin{tabular}{>{\centering}m{0.135\textwidth}
 >{\centering}m{0.135\textwidth}
 >{\centering}m{0.135\textwidth}
 >{\centering}m{0.135\textwidth}
 >{\centering}m{0.135\textwidth}
 >{\centering}m{0.135\textwidth}
 >{\centering\arraybackslash}m{0.135\textwidth}}
 {Original} & \multicolumn{6}{c}{Attacked Spectrograms} \\ \cmidrule(lr){2-7} 
{Spectrograms} & {FGSM} & {DeepFool} & {BIM-a} & {BIM-b} & {JSMA} & {CWA} \\
 \end{tabular}
\centering
 \subfloat[MFCC]{{\includegraphics[width=0.135\textwidth]{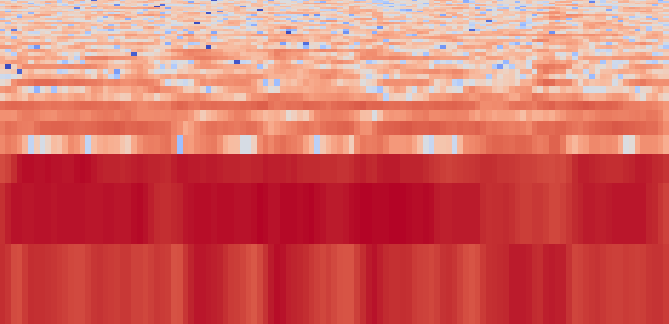} }}%
 \subfloat[$\left \| \delta \right \|_{2}=0.51, l{}'=2$]{{\includegraphics[width=0.135\textwidth]{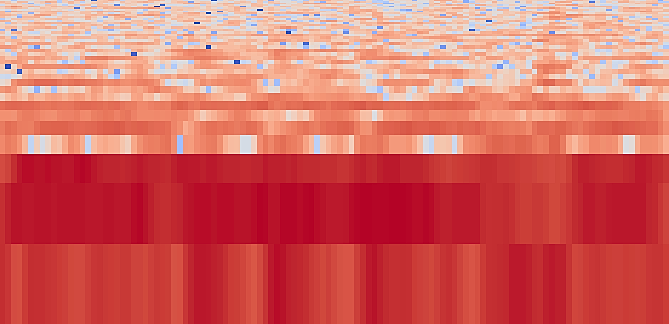} }}
 \subfloat[$\left \| \delta \right \|_{2}=0.67, l{}'=3$]{{\includegraphics[width=0.135\textwidth]{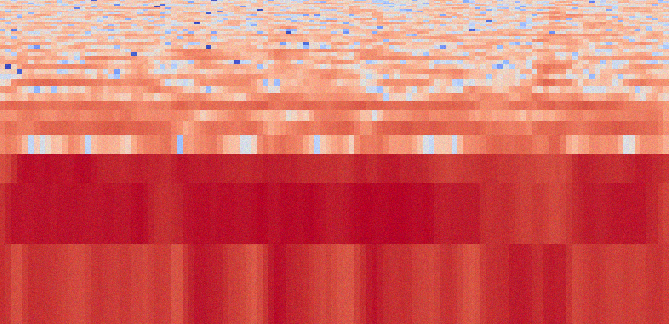} }}%
 \subfloat[$\left \| \delta \right \|_{2}=0.71, l{}'=4$]{{\includegraphics[width=0.135\textwidth]{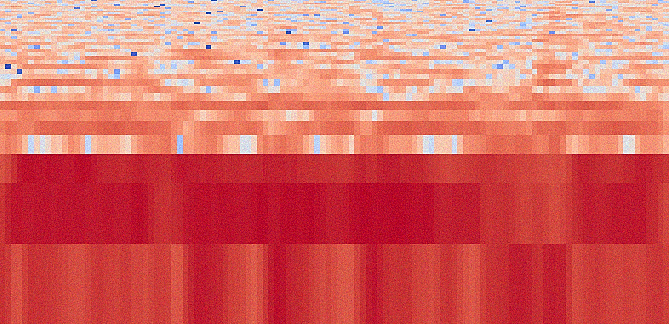} }}%
 \subfloat[$\left \| \delta \right \|_{2}=0.93, l{}'=5$]{{\includegraphics[width=0.135\textwidth]{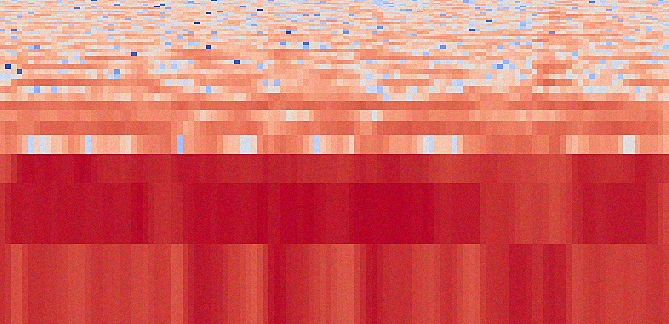} }}%
 \subfloat[$\left \| \delta \right \|_{0}=1.18, l{}'=6$]{{\includegraphics[width=0.135\textwidth]{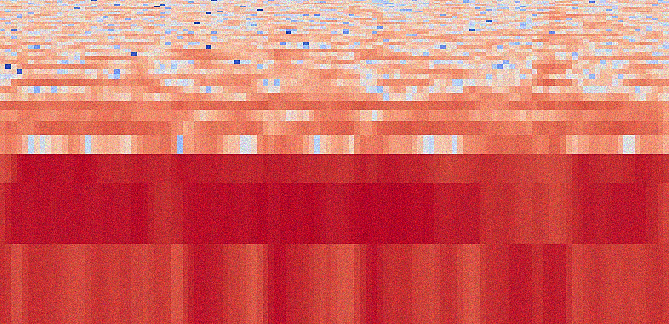} }}%
 \subfloat[$\left \| \delta \right \|_{2}=1.47, l{}'=7$]{{\includegraphics[width=0.135\textwidth]{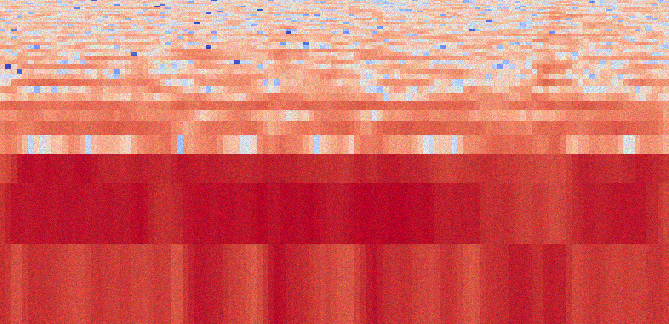} }}%
 \\
 \subfloat[STFT]{{\includegraphics[width=0.135\textwidth]{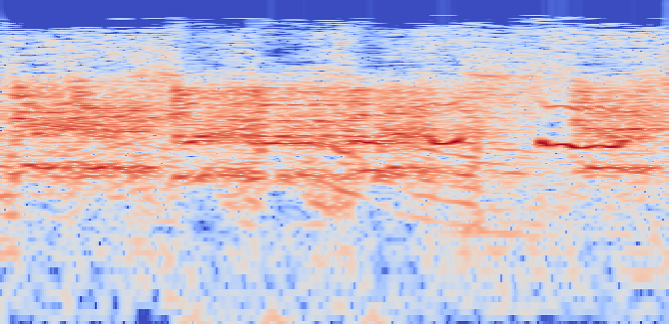} }}%
 \subfloat[$\left \| \delta \right \|_{2}=0.82, l{}'=2$]{{\includegraphics[width=0.135\textwidth]{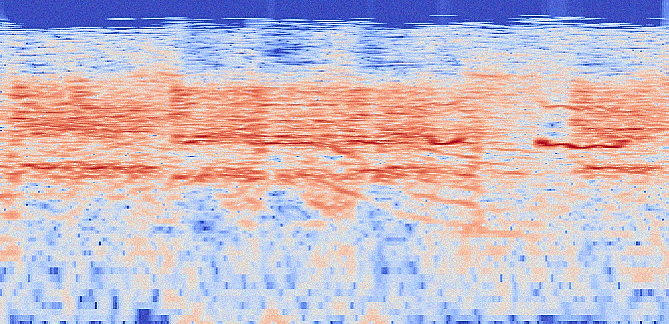} }}%
 \subfloat[$\left \| \delta \right \|_{2}=1.39, l{}'=3$]{{\includegraphics[width=0.135\textwidth]{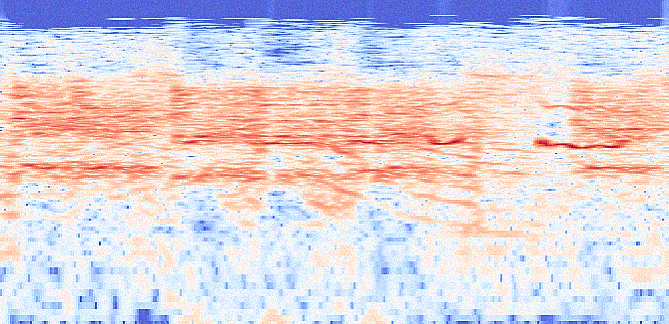} }}%
 \subfloat[$\left \| \delta \right \|_{2}=0.64, l{}'=4$]{{\includegraphics[width=0.135\textwidth]{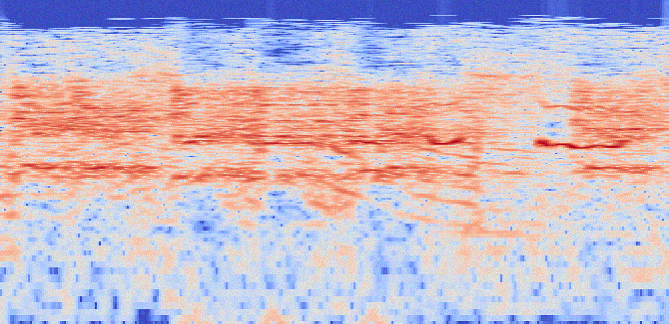} }}%
 \subfloat[$\left \| \delta \right \|_{2}=1.24, l{}'=5$]{{\includegraphics[width=0.135\textwidth]{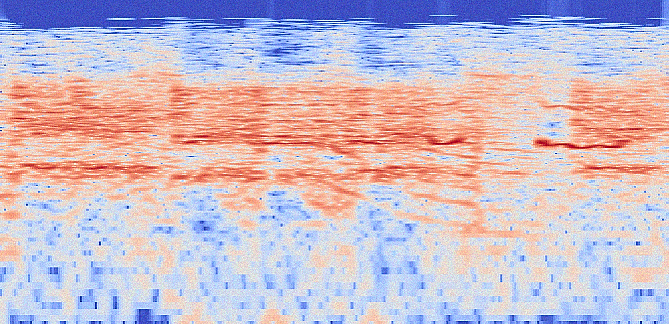} }}%
 \subfloat[$\left \| \delta \right \|_{0}=1.31, l{}'=6$]{{\includegraphics[width=0.135\textwidth]{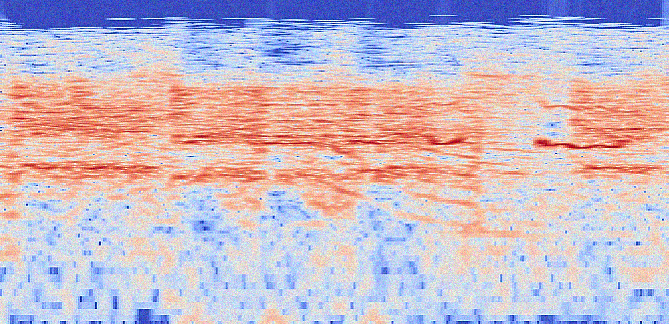} }}%
 \subfloat[$\left \| \delta \right \|_{2}=1.73, l{}'=7$]{{\includegraphics[width=0.135\textwidth]{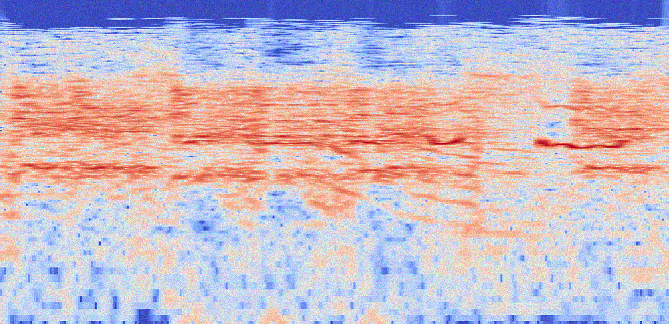} }}%
 \\
 \subfloat[DWT]{{\includegraphics[width=0.135\textwidth]{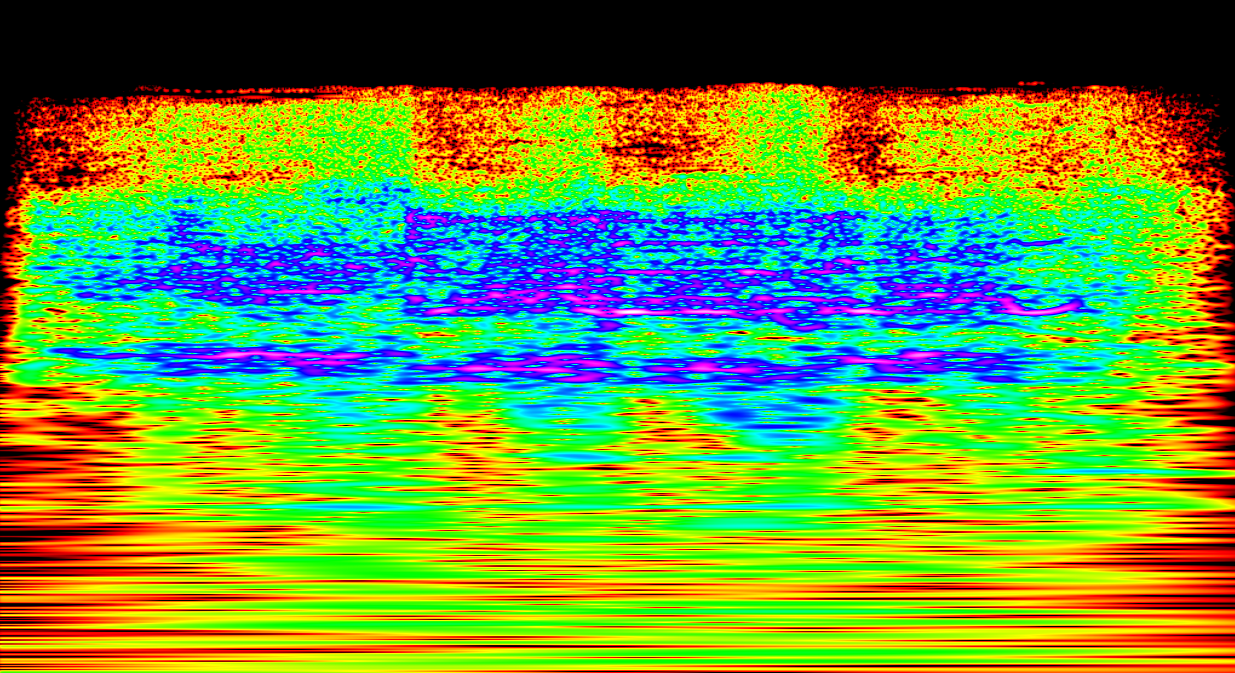} }}%
 \subfloat[$\left \| \delta \right \|_{2}=1.13, l{}'=2$]{{\includegraphics[width=0.135\textwidth]{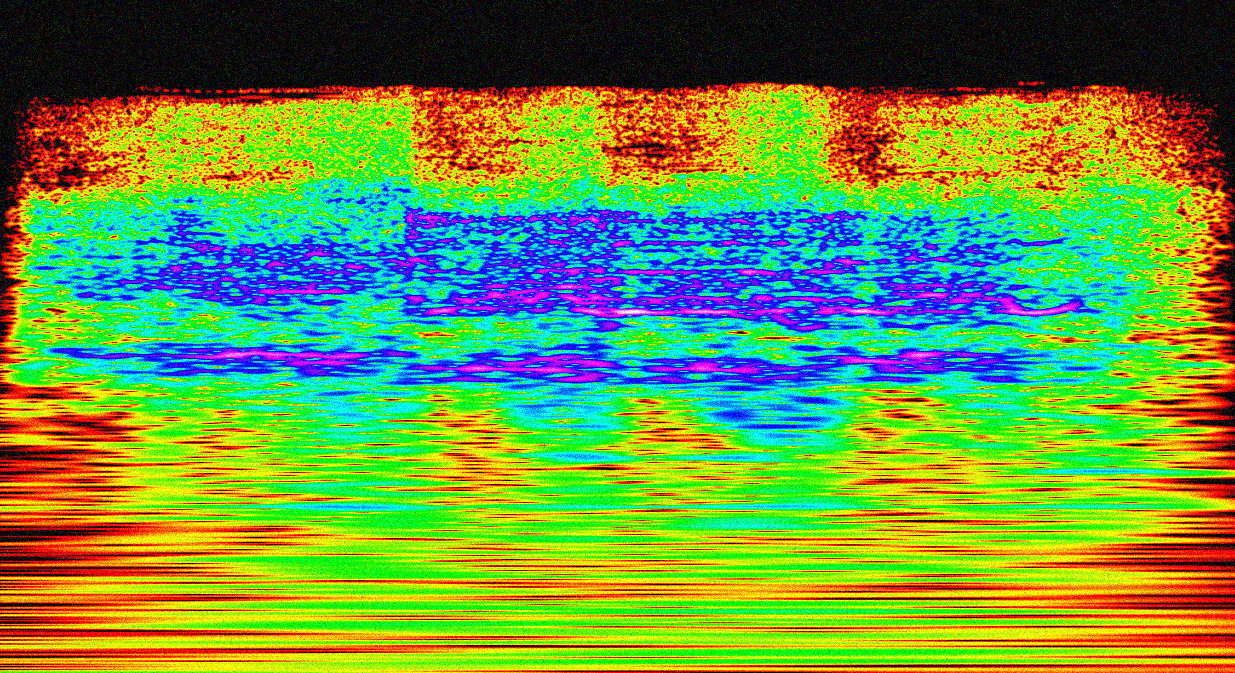} }}%
 \subfloat[$\left \| \delta \right \|_{2}=1.36, l{}'=3$]{{\includegraphics[width=0.135\textwidth]{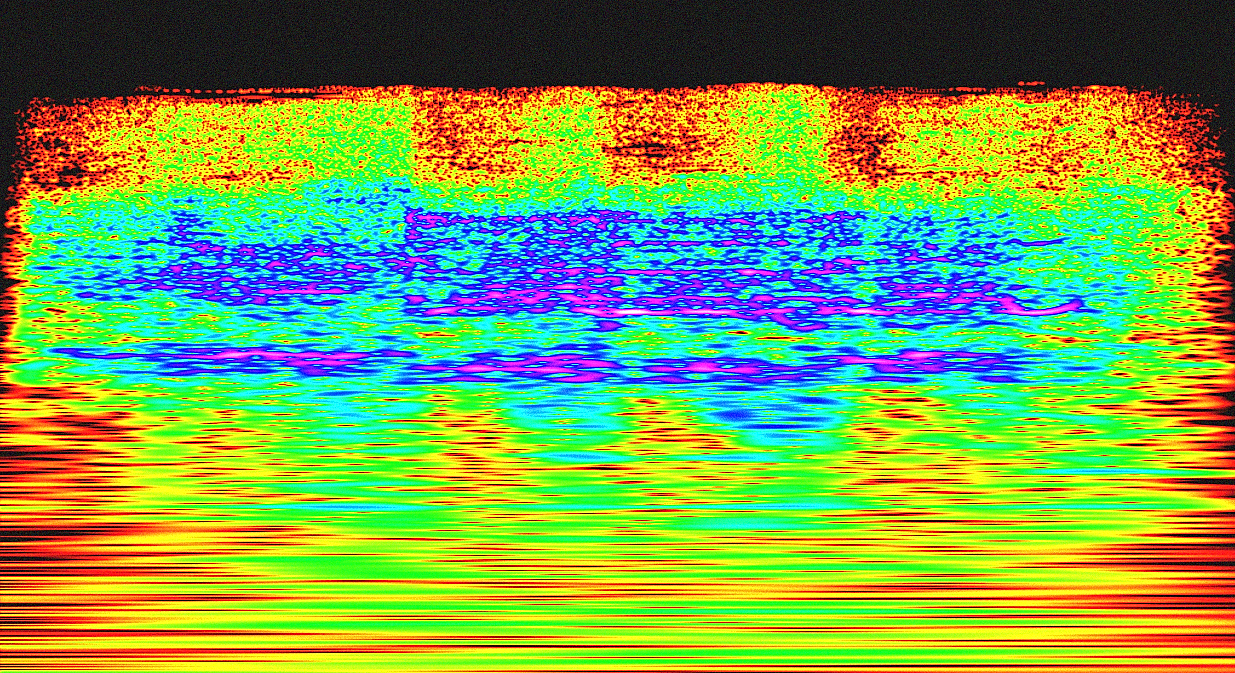} }}%
 \subfloat[$\left \| \delta \right \|_{2}=1.96, l{}'=4$]{{\includegraphics[width=0.135\textwidth]{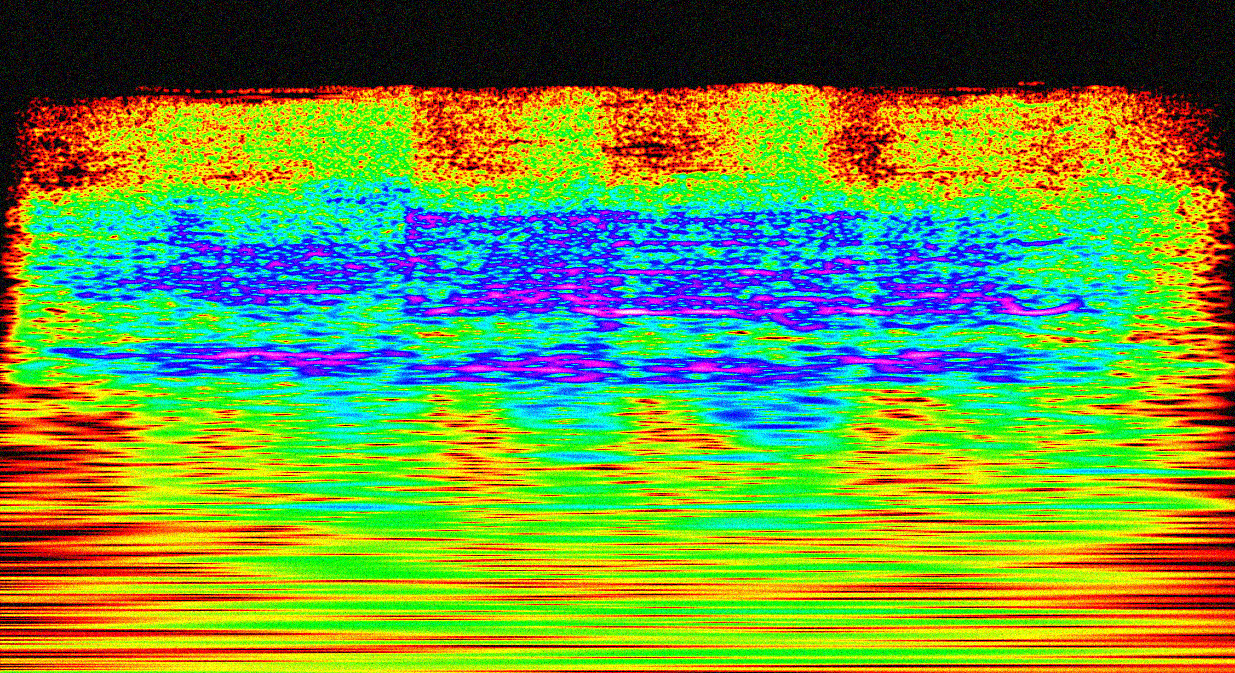} }}%
 \subfloat[$\left \| \delta \right \|_{2}=1.49, l{}'=5$]{{\includegraphics[width=0.135\textwidth]{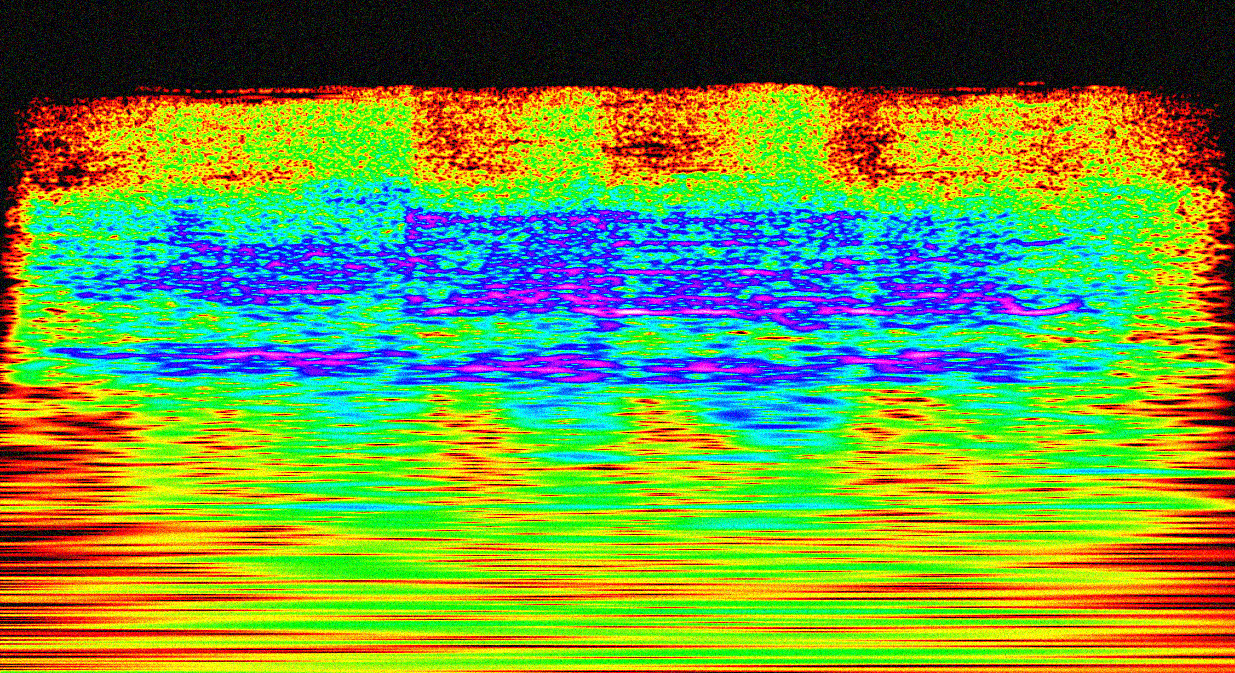} }}%
 \subfloat[$\left \| \delta \right \|_{0}=2.03, l{}'=6$]{{\includegraphics[width=0.135\textwidth]{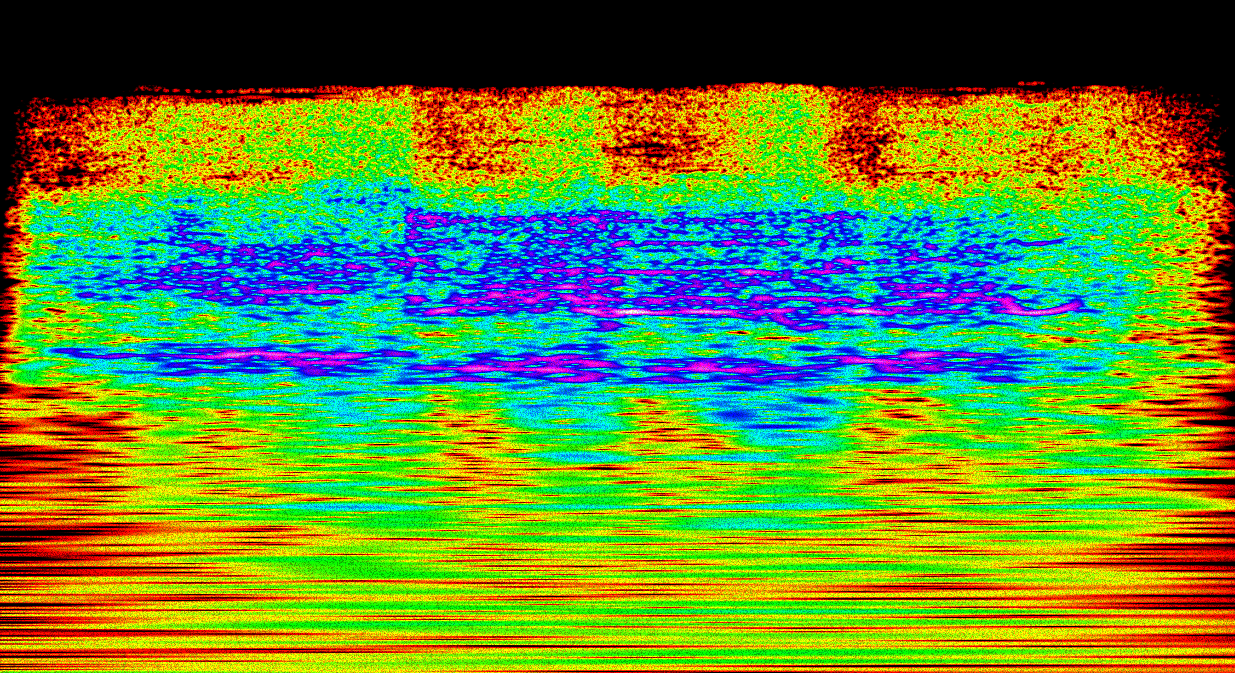} }}%
 \subfloat[$\left \| \delta \right \|_{2}=2.38, l{}'=7$]{{\includegraphics[width=0.135\textwidth]{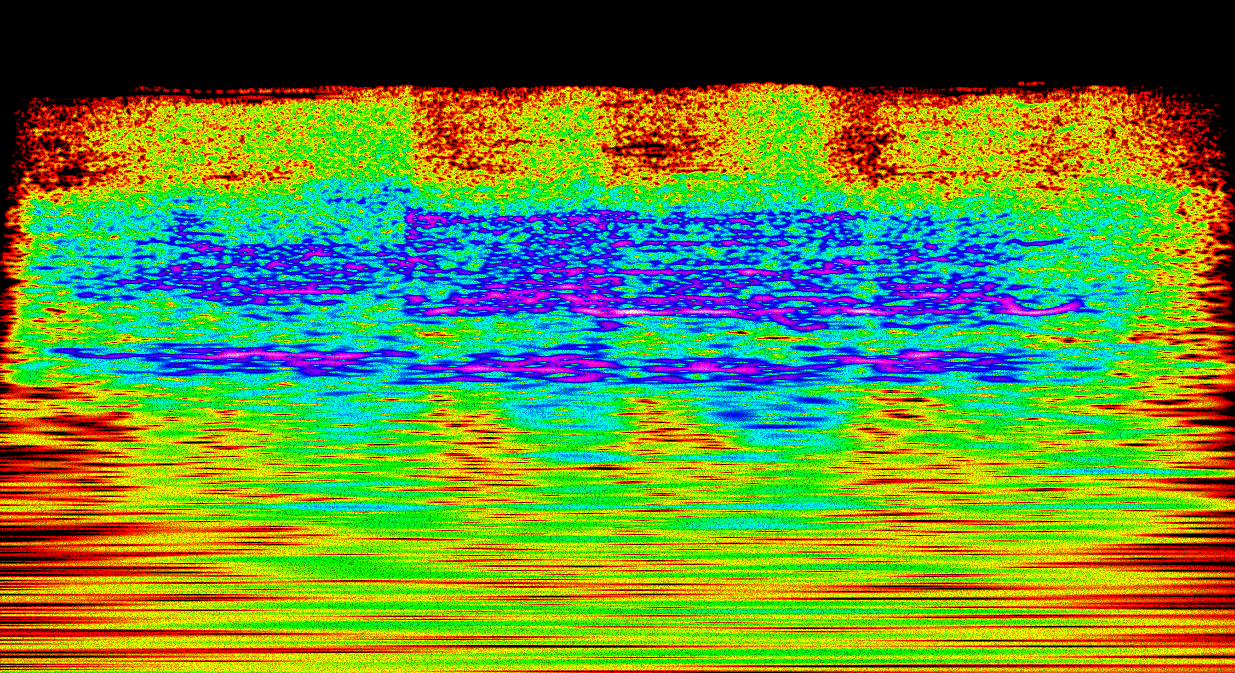} }}%
\caption{Crafted adversarial spectrograms for the three audio representations. The original audio sample has been randomly selected from the class of dog bark ($l=1$). Examples shown in columns two to seven are associated with the six adversarial attacks for the original input sample. Required perturbation ($\delta$) and the target labels ($l{}'$) are shown under each spectrogram.}
\label{attack_spec_compr}
\end{sidewaysfigure}

In all these experiments, we assumed a frame length of 50 ms with 50\% overlapping to convolve the input signal with mother functions. We have also carried out experiments on studying the potential effect of frame length on the performance of the models. They showed that short frame lengths (e.g., 30 ms) drop the recognition performance of the models for the three benchmark datasets. Additionally, short and very long frames such as 30 ms and 70 ms introduce insufficient overlap in frequency plots, which result in dropping the recognition accuracy (see Figure~\ref{mel-attack-pack5}). Figure~\ref{attack_spec_compr} visually compares crafted adversarial examples for the three representations. Although they are visually very similar to their legitimate counterparts, they confidently drive the classifier toward wrong predictions. That showcases the active threat of adversarial attacks for the sound recognition models.

\captionsetup[subfigure]{labelformat=empty}

\section{Discussion}
\label{discuss:sec}
In this section, we provide additional discussion regarding our results. We briefly discuss some secondary aspects of our experiments that could be relevant for future studies. 

\subsection{Deep Learning Architectures} 
We measured recognition accuracy and the total number of training parameters for all candidates for selecting the front-end classifier. We explored DL architectures without residual blocks (AlexNet) and with inception blocks (GoogLeNet) to choose victim classifiers. Table~\ref{table:compRelDif} unveils that these dense networks do not outperform ResNet-18 in terms of recognition accuracy. Although the average recognition accuracy of ResNet-18 and GoogLeNet are competitive on spectrograms, the latter has 1.41$\times$ more training parameters. On average, the recognition performance of AlexNet is 8\% lower than ResNet-18, even if it has 61\% fewer parameters. Furthermore, the recognition performance of other ResNet models such as ResNet-34 and ResNet-56 are very competitive to ResNet-18, but the latter requires 50\% fewer parameters. 
\textcolor{black}{Table~\ref{table:compRelDif} also shows the performance of two additional sound recognition architectures, namely SB-CNN \cite{salamon2017deep} and the long short-term memory (LSTM) \cite{das2020urban}, which have been widely benchmarked both for classification and adversarial attack studies \cite{esmaeilpour2021towards}. The first classifier employs a convolutional network for extracting local features from a MFCC representation, and it relatively requires fewer training parameters than ResNet-18. On the contrary, the LSTM-based classifier enquires considerably higher training parameters since it exploits multiple gates configuration for modeling small frequency variations for the signal.} 

In comparing the robustness of all these models against adversarial attacks, they can reach fooling rates higher than 95\%. Considering the allocated budgets, the ResNet-18 is the costliest network in terms of the number of required gradient computations for the adversary, followed by GoogLeNet, AlexNet, \textcolor{black}{LSTM-based and SB-CNN,} as shown in Table~\ref{table:compRelDif}.
\begin{table}[t]
\scriptsize
\setlength{\tabcolsep}{0.5em}
\centering
\caption{Comparison of some front-end classifiers in terms of average recognition accuracy, fooling rate, the ratio of the training parameters, and the cost of attack (gradient computation) relative to ResNet-18. Herein, $+$ $\uparrow$ and $-$ $\downarrow$ symbols denote the relative increase and decrease of the comparison metric, respectively.}
\begin{tabular}{|c||c|c|c|c|c|c|}
\hline
                                                           & \multicolumn{6}{c|}{{{CNN Architecture}}} \\ \cline{2-7}                 \multicolumn{1}{|c||}{{{Comparison Metric}}}   & {{ResNet-34}} & {{ResNet-56}} & {{AlexNet}}   & {{GoogLeNet}} & \textcolor{black}{SB-CNN \cite{salamon2017deep}} & \textcolor{black}{LSTM \cite{das2020urban}}\\ \hline \hline
{{Difference in Recognition Accuracy (\%)}} & $ -02.78\downarrow$   & $ - 01.92\downarrow$  & $ - 08.16\downarrow$  & $ - 06.58\downarrow$ & $\color{blue} -09.52 \downarrow$ &  $\color{blue} -02.27 \downarrow$\\ \hline

{{Difference in Fooling Rate (AUC Score)}} & $ -03.14\downarrow$   & $ - 04.34\downarrow$  & $ - 12.09\downarrow$  & $ - 08.63\downarrow$ & $\color{blue} -02.18 \downarrow$ & $\color{blue} - 01.15 \downarrow$  \\ \hline
{{Training Parameter Ratio  (\%)}}                                                & $ + 51.11\uparrow$ & $ +49.88\uparrow$  & $  + 61.32\uparrow$ & $ + 40.26\uparrow$ & $\color{blue} - 14.14 \downarrow$ & $\color{blue} + 36.39 \uparrow$ \\ \hline
{{Gradient Computation Ratio (\%)}}                                                      & $ - 43.17\downarrow$ & $ - 27.71\downarrow$ & $ - 31.40\downarrow$ & $ - 21.09\downarrow$ & $\color{blue} -51.22 \downarrow$ & $\color{blue} -31.44 \downarrow$ \\ \hline
\end{tabular}
\label{table:compRelDif}
\end{table}

\subsection{Data Augmentation} 
To improve the classifiers' performance, we augmented the original datasets only at waveform level (1D) using time-stretching filter except for DWT representations which we additionally scaled the spectrograms by a logarithmic function. Removing 1D data augmentation negatively affects recognition accuracy of the models with drop ratios of about 0.056\%, 0.036\%, and 0.029\% for MFCC, STFT, and DWT spectrograms, respectively. To measure these models' robustness against adversarial examples, we executed attack algorithms on random batches of size 100 among the entire datasets. The experimental results have shown that for reaching the fooling rates as close as the values reported in Tables~\ref{mfcc_sr_effect} to \ref{dwt_effect}, less gradient computation is required mainly for JSMA and CWA attacks.

\subsection{{Brief Insight over DWT Representation}} 
As mentioned in Section~\ref{audio_rep_section}, DWT employs different mother functions which can provide an extensive range of local spectral features (from low to high resolutions) \cite{pathak2009wavelet}. Compared to the sinusoidal basis functions in the Fourier-based representations (i.e., MFCC and STFT), these functions are considerably complex. However, they help extract more discriminative features from the audio signals and consequently improve the learning performance of the classifier~\cite{stephane1999wavelet}. This is experimentally shown in Tables 2, 3, and 4, where we have  demonstrated that our front-end classifiers trained on DWT spectrograms outperform the models trained on other 2D representations.


\subsection{{The Impact of Mel-Spectrogram Settings on the Performance of the Adversarial Attacks}}
\label{sec:impactMelSPectroPer}
{
Mel-spectrogram is another signal representation approach, which is very similar to the STFT spectrogram. However, it employs a scaling policy for expanding the short-term Fourier spectrum of a signal over a nonlinear distribution. This policy incorporates a set of predefined basis functions for transforming frequency coefficients into Mel scales. The motivation behind developing such a representation is extracting audio features, which are more correlated to the human auditory system~\cite{hershey2017cnn,dong2018convolutional,ren2019fastspeech}.
}

{
For investigating the potential relation between the Mel-spectrogram settings and the fooling rate of the adversarial attacks, we use the Librosa toolkit library and follow the similar settings as discussed in Section~\ref{sec:advAttForSTFTPB} and \ref{sec:advAttForSTFTPP}. Table~\ref{melSpec_fft_effect} summarizes the performance of the front-end classifier in terms of recognition accuracy and robustness against six benchmarking adversarial attacks. As expected, the statistics in this table are very close to the values reported in Table~\ref{stft_fft_effect}.}
{
\begin{table}
\centering
\scriptsize
\setlength{\tabcolsep}{0.5em}
\caption{{Performance comparison of models trained on Mel-spectrogram with different $N_{\text{FFT}}$ averaged over experiments and budgets. Relatively better performances are in boldface.}}
\begin{tabular}{|c|c|c||c|c|c|c|c|c|}
\hline
\multirow{2}{*}{Dataset} & Number & Recog. & \multicolumn{6}{c|}{AUC Score for Fooling Rate, Number of Gradients for Adversarial Attacks} \\ \cline{4-9} 
 & of FFTs & Acc. (\%) & FGSM & DeepFool & BIM-a & BIM-b & JSMA & CWA \\ \hline \hline
\multirow{3}{*}{ESC-10} & 512 & {81.53} & {0.9623}, 1 & {0.9343}, {210} & {0.9379}, {088} & {0.9137}, {285} & {0.9409}, {305} & {0.9517}, {1646} \\ \cline{2-9} 
 & 1~024 & \textbf{{84.67}} & \textbf{{0.9801}}, 1 & \textbf{{0.9664}}, {185} & \textbf{{0.9588}}, {281} & \textbf{{0.9450}}, {106} & {\textbf{0.9423}}, {288} & \textbf{{0.9780}}, {1893} \\ \cline{2-9} 
 & 2~048 & {77.03} & {0.9428}, 1 & {0.9410}, {176} & {0.9514}, {127} & {0.9101}, {377} & {0.9356}, {593} & {0.8639}, {2011} \\ \hline \hline
\multirow{3}{*}{ESC-50} & 512 & {82.56} & {0.9572}, 1 & {0.9482}, {116} & {0.9459}, {238} & {0.9586}, {178} & {0.9610}, {199} & {0.9281}, {2119} \\ \cline{2-9} 
 & 1~024 & \textbf{{83.96}} & \textbf{{0.9735}}, 1 & {0.9497}, {272} & \textbf{{0.9742}}, {182} & \textbf{{0.9663}}, {318} & {0.9679}, {314} & \textbf{{0.9705}}, {2675} \\ \cline{2-9} 
 & 2~048 & {79.99} & {0.9416}, 1 & \textbf{{0.9562}}, {307} & {0.9340}, {205} & {0.9452}, {401} & \textbf{{0.9698}}, {436} & {0.9136}, {2237} \\ \hline \hline
\multirow{3}{*}{US8k} & 512 & {86.04} & {0.9688}, 1 & {0.9220}, {411} & \textbf{{0.9377}}, {515} & {0.9315}, {164} & {0.9225}, {208} & {0.9076}, {2758} \\ \cline{2-9} 
 & 1~024 & {89.55} & {0.9772}, 1 & {0.9603}, {223} & {0.9011}, {439} & \textbf{{0.9521}}, {633} & {0.9348}, {451} & {0.9244}, {3099} \\ \cline{2-9} 
 & 2~048 & \textbf{{90.17}} & \textbf{{0.9856}}, 1 & \textbf{{0.9632}}, {591} & {0.9231}, {593} & {0.9773}, {744} & \textbf{{0.9507}}, {663} & \textbf{{0.9302}}, {3521} \\ \hline
\end{tabular}
\label{melSpec_fft_effect}
\end{table}
}

{The experimental results have shown that scaling $N_{\mathrm{FFT}}$ (similar to the experiment presented in Section~\ref{sec:advAttForSTFTPP} and demonstrated in Fig.~\ref{mel-attack-pack4}) in generating Mel-spectrogram does not improve the recognition accuracy of the classifier. Furthermore, it slightly decreased the gradient computations with a ratio of 0.006 relative to the STFT representation and averaged over scales in the range $\left [ 0.2, 1 \right ]$. Our conjecture about these achieved results is that since Mel-scales filter the high-frequency components (those segments of a signal with too much variation) \cite{rao2010fast}, changing the scale of $N_{\mathrm{FFT}}$ (which are mostly related with high-frequency components) does not noticeably affect the Mel-spectrogram's distribution. Moreover, we noticed that scaling $N_{\mathrm{FFT}}$, particularly for scales between 0.3 and 0.5, blurs the resulting spectrograms partially and this negatively affects the performance of the front-end classifier.  
}

\subsection{{The Effect of Spectrogram Colour on the Performance of Classifiers}}


{The intensity or color of each pixel in a spectrogram indicates the amplitude of a particular frequency at a particular time. Therefore, visualizing a spectrogram in different colour spaces or with different colour maps might affect the classifier's performance. For measuring such an impact, we run additional experiments.
}

{In a big picture, there are four main colour maps for visualising any $n$-dimensional matrix (such as 2D spectrogram) into colourful images, namely sequential (or multi-sequential), diverging, cyclic, and qualitative
\cite{crameri2020misuse}. These colour maps encompass a large stack of colour bins relative to human visual understanding. For investigating the effect of these maps on the performance of the classifiers, we run additional experiments. According to our achieved results, none of the abovementioned colour maps could noticeably affect the performance of the victim classifiers, namely ResNet-18, ResNet-34, ResNet-56, AlexNet, GoogLeNet, SB-CNN, and LSTM.}

{We extend the above experiment to investigate the effect of different colour spaces on the performance of the classifiers. Since packages like Librosa employ the false-colour lookup tables policy \cite{prakash1993sacimage,towsey2014visualization} for generating spectrograms,
the only remaining valid colour scale (in terms of making noticeable changes on the performance of the classifier) is the uniformly distributed HSV (hue, saturation, and value) representation. For every benchmarking environmental sound dataset, we generate such representations and follow the same settings as discussed in Sections~\ref{sec:advAttForMFCCPB}, \ref{sec:advAttForSTFTPB}, and \ref{sec:advAttForDWTPB}. The achieved results are summarised in Table~\ref{Table:hsvhsv}.
}
\begin{table}[t]
\centering
\scriptsize
\setlength{\tabcolsep}{0.5em}
\caption{{The comparison of recognition accuracy and fooling rate of the front-end classifiers on the spectrograms visualised with uniformly distributed HSV representation. The recognition accuracy (Acc. \%) and the AUC score associated with the fooling rate (FR) are averaged over different experiments and attack budgets. Herein, Mel-Spec denotes the Mel-Spectrogram mentioned in Section~\ref{sec:impactMelSPectroPer}.}}
\begin{tabular}{|c||cc|cc|cc|cc|}
\hline
         & \multicolumn{2}{c|}{MFCC}           & \multicolumn{2}{c|}{STFT}           & \multicolumn{2}{c|}{DWT}            & \multicolumn{2}{c|}{Mel-Spec}      \\ \cline{2-9} 
Datasets & \multicolumn{1}{c|}{Acc.}  & FR     & \multicolumn{1}{c|}{Acc.}  & FR     & \multicolumn{1}{c|}{Acc.}  & FR     & \multicolumn{1}{c|}{Acc.}  & FR     \\ \hline \hline
ESC-10   & \multicolumn{1}{c|}{{68.36}} & {0.7022} & \multicolumn{1}{c|}{{72.11}} & {0.6906} & \multicolumn{1}{c|}{{67.99}} & {0.6382} & \multicolumn{1}{c|}{{67.29}} & {0.7011} \\ \hline
ESC-50   & \multicolumn{1}{c|}{{64.55}} & {0.7519} & \multicolumn{1}{c|}{{64.59}} & {0.8368} & \multicolumn{1}{c|}{{72.48}} & {0.7888} & \multicolumn{1}{c|}{{61.52}} & {0.6679} \\ \hline
UrbanSound8k     & \multicolumn{1}{c|}{{60.96}} & {0.6148} & \multicolumn{1}{c|}{{71.93}} & {0.7533} & \multicolumn{1}{c|}{{80.16}} & {0.8805} & \multicolumn{1}{c|}{{63.24}} & {0.6911} \\ \hline
\end{tabular}
\label{Table:hsvhsv}
\end{table}

{As shown in Table~\ref{Table:hsvhsv}, the recognition accuracy and the fooling rate of the classifiers trained on the spectrograms visualised with the uniformly distributed HSV colour scale are relatively lower than the models trained on spectrograms visualised with the standard logarithmic RGB scale (relative to Tables~\ref{mfcc_sr_effect}, ~\ref{stft_fft_effect}, ~\ref{dwt_effect}, and~\ref{melSpec_fft_effect}). Moreover, Table~\ref{Table:hsvhsv} shows that the fooling rates of the front-end classifiers trained on the uniformly visualised HSV spectrograms are considerably below the 90\% threshold. Therefore, this colour scale does not help make a reasonable trade-off between high recognition accuracy and fooling rate compared to the baseline logarithmic RGB. This is a critical aspect since our focus is investigating major spectrogram settings (e.g., sampling rate, $N_{\mathrm{FFT}}$, mother function, etc.), which can make a reasonable trade-off between high recognition accuracy and AUC score of the imposed adversarial attacks.}


\subsection{Adversarial on Raw Audio} 
Optimizing Eq.~\ref{general_adv_formula} even for a short 1D audio signal sampled at a low rate is very costly, and they are not transferable while being played over the air \cite{carlini2018audio}. We trained several end-to-end ConvNets on randomly selected batches of environmental sound datasets to address this interesting open problem. Upon running both targeted and non-targeted attacks against ConvNets, we could reduce the performance of victim classifiers by 30\% on average. Interestingly, multiplying the adversarial examples by a small random scalar restored the audio waveforms' correct label. In other words, whereas adversarial spectrograms, 1D adversarial audio waveforms are not resilient against any additional perturbation.

\subsection{Adversarial Transferability} 
The transferability of adversarial examples is not only dependent on the classifier but also on audio representations. We investigated this aspect on deep neural networks trained on different spectrograms. Table~\ref{table_transf} reports the transferability ratios averaged over budgets with batch sizes of 100. Crafted adversarial examples for victim models are less transferable in MFCC representations, while DWT spectrograms have higher transferring rates on average. On the other hand, examples generated in the STFT domain are more transferable compared to MFCC. That may be due to the higher order of information in STFT spectrograms.

\begin{table*}
\centering
\scriptsize
\setlength{\tabcolsep}{0.1em}
\caption{Average transferability ratio of adversarial examples among ConvNets. Higher ratios are shown in boldface.}
\begin{tabular}{|c||c||c|c|c||c|c|c||c|c|c|}
\hline
 & & \multicolumn{3}{c||}{MFCC} & \multicolumn{3}{c||}{STFT} & \multicolumn{3}{c|}{DWT} \\ \cline{3-11} 
\multirow{-2}{*}{Dataset} & \multirow{-2}{*}{Models} & ResNet-18 & GoogLeNet & AlexNet & ResNet-18 & GoogLeNet & AlexNet & ResNet-18 & GoogLeNet & AlexNet \\ \hline
 & ResNet-18 & {\cellcolor[HTML]{EAECEE}1} & \textbf{0.672} & 0.568 & {\cellcolor[HTML]{EAECEE}1} & \textbf{0.713} & 0.641 & {\cellcolor[HTML]{EAECEE}1} & 0.761 & \textbf{0.774} \\ \hhline{|~|-|-|-|-|-|-|-|-|-|-|} 
 & GoogLeNet & \textbf{0.693} & {\cellcolor[HTML]{EAECEE}1} & 0.480 & \textbf{0.637} & {\cellcolor[HTML]{EAECEE}1} & 0.519 & 0.646 & {\cellcolor[HTML]{EAECEE}1} & \textbf{0.684} \\ \hhline{|~|-|-|-|-|-|-|-|-|-|-|} 
\multirow{-3}{*}{ESC-10} & AlexNet & 0.491 & \textbf{0.521} & {\cellcolor[HTML]{EAECEE}1} & 0.540 & \textbf{0.562} & {\cellcolor[HTML]{EAECEE}1} & 0.633 & \textbf{0.701} & {\cellcolor[HTML]{EAECEE}1} \\ \hline \hline
 & ResNet-18 & {\cellcolor[HTML]{EAECEE}1} & \textbf{0.644} & 0.519 & {\cellcolor[HTML]{EAECEE}1} & \textbf{0.661} & 0.609 & {\cellcolor[HTML]{EAECEE}1} & \textbf{0.755} & 0.732 \\ \hhline{|~|-|-|-|-|-|-|-|-|-|-|} 
 & GoogLeNet & \textbf{0.630} & {\cellcolor[HTML]{EAECEE}1} & 0.531 & \textbf{0.578} & {\cellcolor[HTML]{EAECEE}1} & 0.569 & 0.507 & {\cellcolor[HTML]{EAECEE}1} & \textbf{0.676} \\ \hhline{|~|-|-|-|-|-|-|-|-|-|-|}  
\multirow{-3}{*}{ESC-50} & AlexNet & 0.523 & \textbf{0.536} & {\cellcolor[HTML]{EAECEE}1} & 0.551 & \textbf{0.601} & {\cellcolor[HTML]{EAECEE}1} & 0.614 & \textbf{0.699} & {\cellcolor[HTML]{EAECEE}1} \\ \hline \hline
 & ResNet-18 & {\cellcolor[HTML]{EAECEE}1} & 0.627 & \textbf{0.677} & {\cellcolor[HTML]{EAECEE}1} & 0.611 & \textbf{0.710} & {\cellcolor[HTML]{EAECEE}1} & \textbf{0.714} & 0.713 \\ \hhline{|~|-|-|-|-|-|-|-|-|-|-|} 
 & GoogLeNet & \textbf{0.634} & {\cellcolor[HTML]{EAECEE}1} & 0.503 & 0.563 & {\cellcolor[HTML]{EAECEE}1} & \textbf{0.699} & \textbf{0.723} & {\cellcolor[HTML]{EAECEE}1} & 0.707 \\ \hhline{|~|-|-|-|-|-|-|-|-|-|-|} 
\multirow{-3}{*}{US8k} & AlexNet & 0.577 & \textbf{0.583} & {\cellcolor[HTML]{EAECEE}1} & 0.703 & \textbf{0.735} & {\cellcolor[HTML]{EAECEE}1} & 0.705 & \textbf{0.678} & {\cellcolor[HTML]{EAECEE}1} \\ \hline
\end{tabular}
\label{table_transf}
\end{table*}

Unlike other research works~\cite{esmaeilpour2019robust} that have evaluated adversarial transferability among different classifiers to identify the most reliable model considering a black-box attack scenario, we have carried out the transferability experiment to determine the most reliable 2D representation. Therefore, we characterize the impact of 2D representation on the transferability of attacks among different models. In other words, we have demonstrated that classifiers trained on MFCC representations have a lower adversarial transferability ratio than models trained on STFT and DWT.

\subsection{Selection of Benchmarking Adversarial Attacks}
All the attack algorithms evaluated in this paper are comprehensive and still top-notch approaches in generating adversarial examples. Moreover, they are standard benchmarking approaches in developing defense algorithms since they have a unique objective and technique in finding the most fitting adversarial perturbation. See a relevant discussion in \cite{jang2019adversarial,yu2019new}.

\section{Conclusion}
\label{sec:con}
\textcolor{black}{In this paper, we have demonstrated that spectrogram settings such as sampling rate, number of $\mathrm{FFT}$s, frame length, etc., affect both the recognition accuracy and robustness of the victim front-end classifier (i.e., ResNet-18) against adversarial attacks. Furthermore, we characterized an inverse relationship between recognition accuracy and robustness of ResNet-18 trained on 2D representations of environmental audio signals averaged over the allocated budgets by the adversary. This relation is generalizable to other DL architectures (e.g., ResNet-34, ResNet-56, AlexNet, GoogLeNet, SB-CNN, and LSTM). Additionally, we showed that our front-end classifier could reach a very high recognition accuracy when trained on DWT representation. On average, attacking such a model is more costly for the adversary than classifiers trained on MFCC and STFT representations. Moreover, we have examined the transferability of crafted adversarial examples among AlexNet, GoogLeNet, and ResNet-18 for the three spectrogram representations. Our experimental results showed that MFCC spectrograms achieved the lowest transferability ratio, averaged over six different adversarial attacks. Our future studies will investigate this property for networks trained on speech datasets.
}


\section*{Acknowledgment}
This work was funded by the Natural Sciences and Engineering Research Council of Canada (NSERC) under grants RGPIN 2016-04855 and RGPIN 2016-06628.

\bibliography{mybibfile}

\end{document}